\documentclass[prx,tightenlines,notitlepage,aps,epsf,twocolumn,showpacs,superscriptaddress,longbibliography,nofootinbib]{revtex4-2}
\usepackage{graphicx}
\usepackage{booktabs,multirow}
\usepackage{stmaryrd}
\usepackage{booktabs}
\usepackage[dvipsnames]{xcolor}
\usepackage{latexsym,amsmath,mathtools,amssymb,bbm,array,comment,quantikz}
\usepackage[normalem]{ulem}
\usepackage[colorlinks=true,allcolors=blue]{hyperref}

\renewcommand{\(}{\left(}

\renewcommand{\[}{\left[}

\usepackage{titlesec}
\titleformat{\paragraph}[runin]
  {\bfseries} 
  {} 
  {0pt} 
  {} 
\titlespacing{\paragraph}{0pt}{4pt}{2pt}

\begin{document}

\title{Quantum error correction well below the color code threshold}

\author{\mbox{Ali Lavasani}}
\affiliation{ Quantinuum, Broomfield, CO 80021, USA}
\author{\mbox{Adam P. Reed}}
\affiliation{ Quantinuum, Broomfield, CO 80021, USA}
\author{\mbox{Joan M. Dreiling}}
\affiliation{ Quantinuum, Broomfield, CO 80021, USA}
\author{\mbox{Annie J. Park}}
\affiliation{ Quantinuum, Broomfield, CO 80021, USA}
\author{\mbox{Michael Mills}}
\affiliation{ Quantinuum, Broomfield, CO 80021, USA}
\author{\mbox{Matteo Puviani}}
\affiliation{ Quantinuum, Partnership House, Carlisle Place, London, SW1P 1BX, UK}
\author{\mbox{Gabriel Matos}}
\affiliation{ Quantinuum, Partnership House, Carlisle Place, London, SW1P 1BX, UK}
\author{\mbox{M.S. Allman}}
\affiliation{ Quantinuum, Broomfield, CO 80021, USA}
\author{\mbox{Noah Berthusen}}
\affiliation{ Quantinuum, Broomfield, CO 80021, USA}
\author{\mbox{Natalie C. Brown}}
\affiliation{ Quantinuum, Broomfield, CO 80021, USA}
\author{\mbox{Victor E. Colussi}}
\affiliation{ Quantinuum, Broomfield, CO 80021, USA}
\author{\mbox{Brian Estey}}
\affiliation{ Quantinuum, Broomfield, CO 80021, USA}
\author{\mbox{Cameron Foltz}}
\affiliation{ Quantinuum, Broomfield, CO 80021, USA}
\author{\mbox{Daniel Gresh}}
\affiliation{ Quantinuum, Broomfield, CO 80021, USA}
\author{\mbox{Reza Haghshenas}}
\affiliation{ Quantinuum, Broomfield, CO 80021, USA}
\author{\mbox{Akhil Isanaka}}
\affiliation{ Quantinuum, Broomfield, CO 80021, USA}
\author{\mbox{Nikhil Kotibhaskar}}
\affiliation{ Quantinuum, Partnership House, Carlisle Place, London, SW1P 1BX, UK}
\author{\mbox{Anthony Ransford}}
\affiliation{ Quantinuum, Broomfield, CO 80021, USA}
\author{\mbox{Ciarán Ryan-Anderson}}
\affiliation{ Quantinuum, Broomfield, CO 
80021, USA}
\author{\mbox{Brian Neyenhuis}}
\affiliation{ Quantinuum, Broomfield, CO 80021, USA}
\author{\mbox{David Hayes}}
\affiliation{ Quantinuum, Broomfield, CO 80021, USA}
\author{\mbox{Tzvetan S. Metodi}}
\affiliation{ Quantinuum, Broomfield, CO 80021, USA}
\author{\mbox{Andrew C. Potter}}
\affiliation{ Quantinuum, Broomfield, CO 80021, USA}

\begin{abstract}
    Quantum error correction enables scalable fault-tolerant quantum computation by suppressing logical errors exponentially with code distance once physical error rates fall below a critical threshold.
    Here we demonstrate below-threshold operation of triangular color codes with distances $d=3,5,$ and $7$ on Quantinuum's Helios trapped-ion quantum processor.
    Leveraging both circuit- and hardware- level leakage reduction schemes, we achieve logical error rates as low as $5_{-2}^{+4}\times 10^{-6}$, and demonstrate error suppression by over an order of magnitude, $\Lambda=12_{-2}^{+3}$, with each successive increase in code distance. Moreover, we demonstrate below threshold scaling of error rates for Clifford two-qubit randomized benchmarking circuits when encoded in $d=3,5$ color codes, achieving logical error rate of $7^{+3}_{-3}\times 10^{-5}$ per logical two-qubit Clifford at distance $5$ with a suppression factor of $\Lambda=9_{-3}^{+8}$.
    We compare logical performance for existing near-optimal algorithmic decoders, and both convolutional neural network and transformer-based neural network decoders capable of real-time decoding an order of magnitude faster than the syndrome extraction rate.
    These results approach the logical error rates required for megaquop-scale quantum computation and indicate that scaled trapped-ion processors could reach application-relevant performance with substantially smaller code blocks than indicated by previous results in other qubit modalities.
\end{abstract}
\maketitle
Quantum error correction (QEC) encodes quantum information redundantly across multiple physical qubits so a few physical component failures in the circuit would not cause the entire computation to fail. However, this redundancy introduces additional opportunities for failure and increases the expected number of physical component failures. Thus QEC involves a fundamental trade-off: greater redundancy allows more errors to be corrected while simultaneously increasing the expected number of errors. According to the threshold theorem~\cite{knill1998resilient,aharonov1999fault,kitaev1997quantum} if $p$, the probability that each physical component fails, is less than a threshold value $p_\text{th}$, then increasing the redundancy improves the error correction faster than it introduces errors. For topological quantum error correcting codes, this is manifested in the distance scaling formula
$p_L\propto \left(p/p_\text{th}\right)^{\frac{d+1}{2}}$,
where $d$ is the code distance and $p_L$ is the failure probability of the encoded logical circuit~\cite{dennis2001topological,kitaev2003fault,fowler2012surface,he2025composable}. 
The ratio $\Lambda\equiv\frac{p_\text{th}}{p}$ quantifies how strongly $p_L$ gets suppressed when increasing the code distance, and is a measure of how far the physical error rate lies below the threshold.

In this work we demonstrate below threshold distance scaling for memory experiments and Clifford two-qubit randomized benchmarking circuits encoded in the triangular color code~\cite{steane1996error, bombin2006topological,katzgraber2009error,  landahl2011fault,bombin2012strong, bombin2015gauge, kubica2015universal, Lacroix_2025} on the Quantinuum's Helios-1 trapped-ion quantum processor (QPU)~\cite{ransford2026}. We achieve logical error rates of $5.5_{-0.4}^{+0.4}\times 10^{-4}$, $3.8_{-1}^{+1}\times 10^{-5}$ and $5_{-2}^{+4}\times 10^{-6}$ per syndrome extraction cycle for color codes of distances $3$, $5$ and $7$ respectively, which corresponds to the suppression factor of $\Lambda^{\text{Mem.}}=12_{-2}^{+3}$ and error rates of $7^{+1}_{-1}\times 10^{-4}$ and $7^{+3}_{-3}\times 10^{-5}$ per logical Clifford gate for color codes of distances $3$ and $5$ respectively, which corresponds to the suppression factor of $\Lambda^{\text{Log.}}_{3/5}=9^{+8}_{-4}$. Unless otherwise specified, 
reported uncertainties denote $1\sigma$ ($68\%$) Wilson confidence intervals. We use the combination of hardware-level leakage repump and circuit level leakage reduction methods to mitigate the effects  of leakage errors. 
Furthermore, we train two complementary neural-network decoders, a convolutional neural network (CNN) and a round-token transformer (RTT) on emulator data and apply them to the hardware data for memory experiments without further fine-tuning. Both models exhibit below-threshold distance scaling and comparable logical performance (Table~\ref{tab:shot_stats}). Their inference latency of few hundred microseconds offers a route to real-time decoding.

\begin{figure*}[t]
    \centering
    \includegraphics[width=\textwidth]{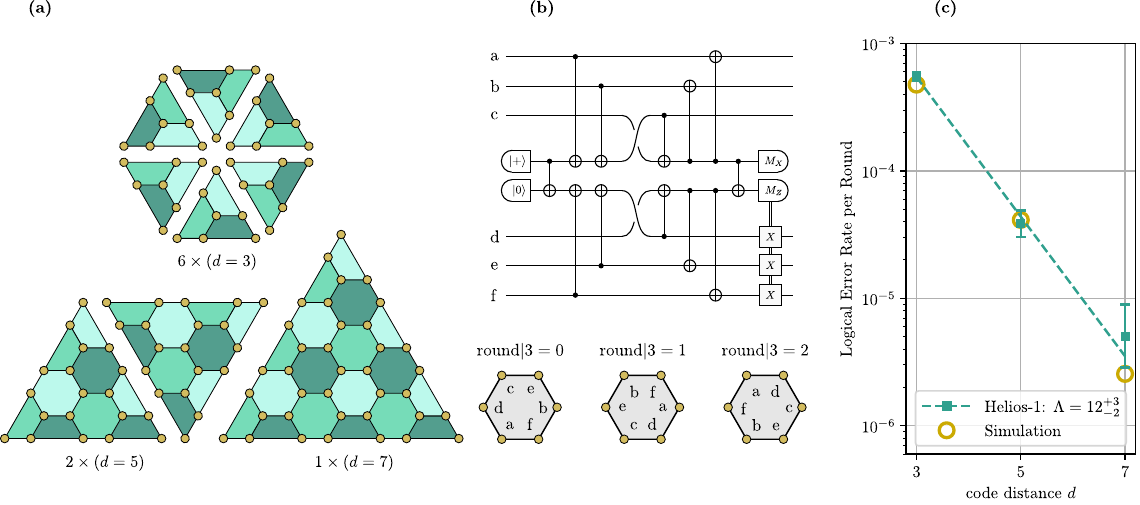}
    \caption{{\bf Triangular color code distance scaling} a) Triangular color codes at distances 3,5 and 7. The gold dots represent physical qubits, and faces correspond to the code stabilizers. For distances $d=3,5$ we run $6$ and $2$ parallel memory experiments per shot. b) The SWAP-superdense syndrome extraction circuit. The bottom figure shows the mapping between the wires in the circuit and the qubits in each face. For four-qubit faces, the missing wires should simply be dropped from the circuit. The mapping changes after each syndrome extraction round, with a period of three, to ensure each data qubits eventually gets swapped with a fresh syndrome qubit. c) The distance scaling plot; both simulation and Helios-1 data points correspond to decoding by Tesseract decoder. The errors for simulation data points are smaller than their markers and hence not shown. }
    \label{fig:main}
\end{figure*}

\noindent\textbf{Triangular Color Code:} The triangular color code family is a self-dual CSS topological quantum error correcting code family with code parameters $\llbracket (3d^2+1)/4, 1, d \rrbracket$ for odd $d$, defined based on the honeycomb tiling of the equilateral triangular surfaces (see Fig.~\ref{fig:main}(a)). The qubits are placed on the vertices, and each face $f$ corresponds to both an $X$-type and a $Z$-type stabilizer generator, given as $\prod_{i\in f} X_i$ and $\prod_{i\in f} Z_i$ respectively. The logical $X$ and $Z$ operators are the product of $X$ and $Z$ on all qubits. Since the code is self-dual, both the logical Hadamard gate $H$ and the logical phase gate $S$ are transversal.  Since it is a CSS code and thanks to the all-to-all connectivity of the ion-trap devices, the logical CNOT gate between two logical code blocks can also be done transversally. Therefore the whole Clifford group can can be done transversally, making this code family particularly interesting for ion-trap platforms.

\begin{figure*}[t]
    \centering

    \begin{minipage}[t]{0.30\textwidth}
        {\raggedright (a)\par}
        
        \includegraphics[width=\linewidth]{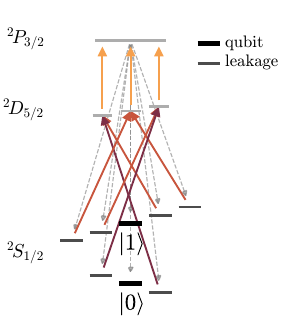}
    \end{minipage}\hspace{\stretch{1}} 
    \begin{minipage}[t]{0.32\textwidth}
        {\raggedright (b)\par}
        \includegraphics[width=\linewidth]{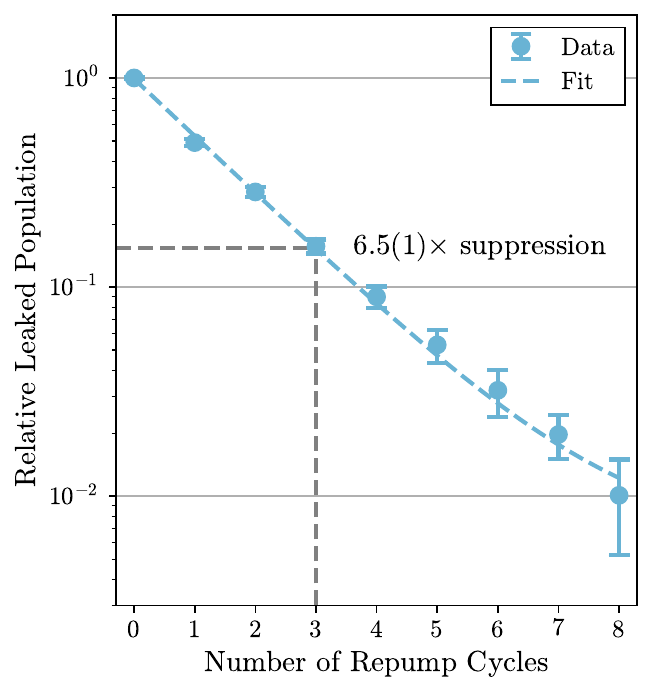}
    \end{minipage}\hspace{\stretch{2}} 
    \begin{minipage}[t]{0.32\textwidth}
        {\raggedright (c)\par}
        \vspace{-5pt}
        \includegraphics[width=\linewidth]{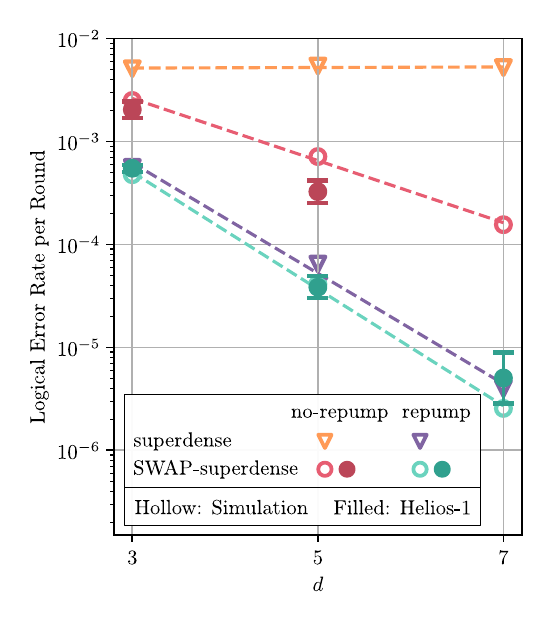}
    \end{minipage}

    \caption{{\bf Hardware-level leakage reduction} a) Leakage repump scheme for $^{137}\mathrm{Ba}^+$. State-selective, narrow-linewidth transitions (red arrows) sequentially transfer leaked population to three ${}^{2}D_{5/2}$ states. After each of these transitions, the transferred population is driven (orange arrows) to ${}^{2}P_{3/2}$, which decays predominantly to ${}^{2}S_{1/2}$ (dashed arrows). b) Leaked population versus number of repump cycles, normalized to the zero-cycle mean (see Methods). Gray dashed lines mark the three cycles used in this work and the corresponding suppression. c) Logical error rate (LER) per syndrome-extraction round, with and without circuit-level LRU and hardware-level leakage repump. Hardware-level leakage repump improves the observed LER by $4\times$ and $8\times$ for $d=3$ and $5$ respectively, compared to the circuit-level LRU alone. 
    }
    \label{fig:repump_overview}
\end{figure*}

\noindent\textbf{Leakage Reduction:}
All leading quantum computing modalities are susceptible to qubits leaking outside their computational subspace~\cite{miao2023, brown2020critical, hrl2026, shreeram2024}, a source of error that can significantly degrade the performance of QEC codes~\cite{fowler2013, brown2019}. Helios QPUs use trapped $^{137}\mathrm{Ba}^+$~ions as qubits, where a physical qubit's computational subspace is defined by a pair of hyperfine levels that lie close in energy to six additional levels into which the qubit can leak, as shown in Fig.~\ref{fig:repump_overview}(a). Leaked population can persist in these levels until a mid-circuit reset returns it to the qubit subspace.
Without mitigation, gates involving leaked qubits spread errors to other qubits, so a single leakage event can trigger a cascade of circuit faults and cause a logical error.

We mitigate leakage using two complementary approaches. First, at the hardware level, we introduce and implement a mid-circuit leakage repump scheme on Helios-1 that optically pumps leaked population back into the qubit subspace. Second, at the circuit level, we use a SWAP-based leakage-reduction unit (LRU) that swaps out data qubits for syndrome extraction ancilla which are then measured and reset~\cite{suchara2015leakage, brown2020critical, mcewen2023relaxing, yoshida2025low}.

The hardware-level scheme repumps leaked population to the computational subspace using the state-selective transitions shown in Fig.~\ref{fig:repump_overview}(a). Within each repump cycle, these transitions are driven sequentially to address all six leakage levels. Because each cycle returns nearly half of the remaining leaked population to the computational subspace, leakage decreases exponentially with the number of cycles~\cite{hayes2020}. We measure the leaked population versus cycle number, shown in Fig.~\ref{fig:repump_overview}(b), and find that the suppression is $6.5(1)\times$ after three cycles, the operating point we use in this work.

At this operating point, the repump scheme adds no time to a circuit and negligible gate error. The repump cycles occur in parallel with ground-state cooling, and both operations overlap with ion transport during qubit sorting~(see Supplement~\ref{sec:repump_conops}). More cycles fit within the cooling time but could increase gate error, so we limit the scheme to three cycles (see Supplement~\ref{sec:repump_benchmarking}). With hardware-level leakage repump, the LER improves at all distances, as shown in Fig.~\ref{fig:repump_overview}(c).

In addition to hardware-level leakage reduction, we use the SWAP-LRU circuit shown in Fig.~\ref{fig:main}(b) to measure the color code stabilizer generators associated to a six qubit face. For faces with four qubits, we simply remove the missing qubits and their corresponding gates. We call this circuit SWAP-superdense, as it is a slight modification of the superdense syndrome extraction circuit~\cite{gidney2023new}. A similar circuit using CXSWAP gates was proposed in Ref.~\cite{yoshida2025low}. The all-to-all connectivity of Helios allows the SWAP gate in the SWAP-superdense circuit to be implemented through software-level qubit relabeling, and hence reduces the syndrome extraction depth by one compared to the original superdense circuit. The classical feedback at the end of the circuit is implemented virtually in software by updating the Pauli frame. 
We measure all the stabilizer generators in parallel, with the scheduling per face shown at the bottom of Fig.~\ref{fig:main}(b).
The syndrome extraction schedule is cycled through three distinct orderings (Fig.~\ref{fig:main}(b)), to ensure that all data qubits get swapped over the course of three syndrome extraction cycles. 
Using this scheduling, qubits in the bulk get swapped at every syndrome extraction round, qubits at the boundary get swapped twice in every three rounds of syndrome extraction and the qubits at the corners get swapped once in every three syndrome extraction cycles. LRU helps with taking care of leaked qubits that stay leaked after hardware-level leakage repump.

\noindent\textbf{Logical Memory Experiment Setup:} 
 The memory experiment we run is as follows: we prepare the code state in one of the six logical Pauli bases $\{\ket{\pm}, \ket{i\pm}, \ket{0},\ket{1}\}$, measure the full set of stabilizer generators for $r=20$ cycles, and then we perform a logical measurement in the appropriate basis at the end. For preparing the logical state, we initialize all qubits in the $\ket{0}$ state and measure all the $X$ stabilizers once using the Bell-flag circuit~\cite{baireuther2019neural} (see Supplementary Material) to prepare the logical $\ket{\bar{0}}$ state and then rotate it to the desired logical state using the required transversal Clifford gate. For logical measure-out, we rotate the logical state back to the $\ket{\bar{0}}$ state and then measure-out all physical qubits. We use the SWAP-superdense circuit shown in Fig.~\ref{fig:main}(b) for syndrome extraction. At the end of the experiment, we feed the syndrome data from the entire circuit to the decoder to decode the value of the logical measurement. The experiment is run for color codes of distance $d=3,5$ and $7$. Distances $d=3,5$ and $7$ require $13$, $37$ and $73$ qubits per memory experiment. Thus, to make the total number of qubits utilized in each shot be almost the same, we run $6$, $2$ and $1$  experiments in parallel in each shot for $d=3,5$ and $7$ respectively(see Fig.~\ref{fig:main}(a)).

{
\setlength{\tabcolsep}{5pt}
\renewcommand{\arraystretch}{1.2} 
\begin{table*} 
\centering 
\caption{{\bf Memory experiment statistics} for increasing code distance $d$. Each shot included $r=20$ rounds of syndrome extractions. The number of failures indicates the number of shots with a logical error detected upon readout. The logical error rate (LER) is reported with 1$\sigma$ (68\%) Wilson confidence intervals. Results are reported for Tesseract~\cite{beni2025tesseract}, Frontier~\cite{leverrier2026}, and the two machine learning decoders trained on the Helios emulator: namely, 3D convolutional neural network (CNN) based and round-token transformer (RTT) based architectures.
}
\label{tab:shot_stats} 
\begin{tabular}{c c cc cc cc cc} \toprule \multirow{2}{*}{$d$} & \multirow{2}{*}{shots (r=20)} & \multicolumn{2}{c}{Tesseract} & \multicolumn{2}{c}{Frontier} & \multicolumn{2}{c}{CNN} & \multicolumn{2}{c}{RTT} \\
\cmidrule(lr){3-4} \cmidrule(lr){5-6} \cmidrule(lr){7-8} \cmidrule(lr){9-10}& & failures & LER & failures & LER & failures & LER & failures & LER \\
\midrule 
3 & 15,840 & 178 & $5.5^{+0.4}_{-0.4}\times10^{-4}$& 166 & $5.1^{+0.4}_{-0.4}\times10^{-4}$ & 171 & $5.3^{+0.4}_{-0.4}\!\times\!10^{-4}$ & 172 & $5.3^{+0.4}_{-0.4}\times10^{-4}$ \\
5 & 21,600 & 17 & $3.8^{+1}_{-0.8}\times10^{-5}$ & 18 & $4.0^{+1}_{-0.9}\times10^{-5}$ & 37 & $8^{+1}_{-1}\!\times\!10^{-5}$ & 29 & $7^{+1}_{-1}\times10^{-5}$ \\
7 & 29,000 & 3 & $5^{+4}_{-2}\times10^{-6}$ & 3 & $5^{+4}_{-2}\times10^{-6}$ & 3 & $5^{+4}_{-2}\!\times\!10^{-6}$ & 5 & $8^{+5}_{-3}\times10^{-6}$ \\ \bottomrule \end{tabular}
\end{table*}
}
\noindent\textbf{Decoding:}
\begin{figure}[b]
    \centering
    \includegraphics[width=\columnwidth]{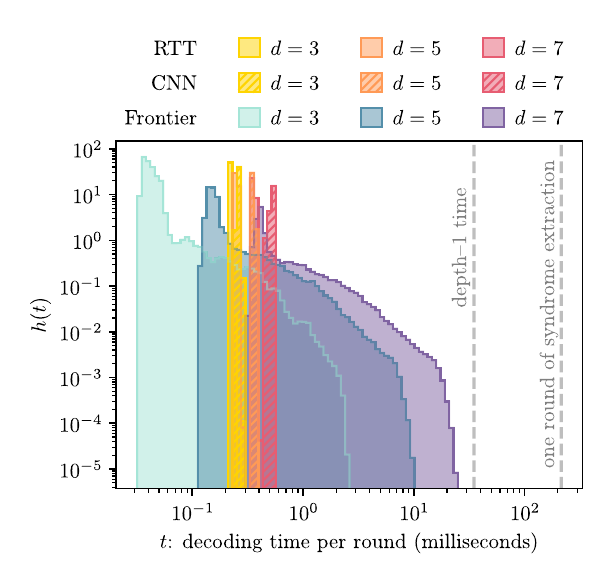}
    \caption{{\bf Decoder statistics} 
    Decoding time distributions for streaming Frontier, CNN and RTT decoders across all three code distances; $h(t)\, \text{d}t$ corresponds to the fraction of rounds, across all shots of a given distance, that took between $t$ and $t+\text{d}t$ milliseconds to process. The numbers for the streaming Frontier is from its WASM implementation. 
    }
    \label{fig:decoding_stats}
\end{figure}
Throughout this work, we use Tesseract~\cite{beni2025tesseractdecoder},  an $A^\ast$-search based most likely error decoder, as our primary decoder. While Tesseract achives near optimal performance, it is generally not fast enough to be used in real time. Hence, to explore potential near optimal real-time decoding solutions, we also consider three streaming decoders for the memory experiment: a streaming version of Frontier~\cite{leverrier2026}, which is a pruned dynamic-programming decoder, a 3D convolutional neural network (CNN), and a round-token transformer (RTT), named this way because it encodes each syndrome extraction round into a token. The Tesseract and Frontier decoders, are configured with a detector error model (DEM) built using Stim~\cite{gidney2021stim}, based on a simplified stochastic Pauli error model of the Helios error channels.
Both neural-network decoders are trained solely on data generated by the Helios-1 emulator, running the same circuits that were run on hardware, to capture detailed hardware noise without relying on a DEM. They are applied to hardware data without further training or fine-tuning. The CNN learns local space--time fault patterns through convolutions; the transformer encodes each round into a token, a vector of numerical features learned from the syndrome data, and learns temporal correlations through an attention mechanism~\cite{vaswani_2017}. 
The methods section provides further details on all three decoders; neural net architecture (Figs.~\ref{fig:CNN3Ddecoder} and~\ref{fig:MLdecoder}) and training details are provided in the Supplementary Material.

All three streaming decoders support processing syndrome data as they arrive. Therefore, while all decoding in this work has been done offline, the syndromes have been fed to the streaming decoders one round at a time to study their processing time that is relevant for real time decoding. 
The Frontier decoding algorithm is naturally amenable to be used in the streaming mode (see Methods).   For the CNN, online predictions reuse cached activations---the intermediate numerical values computed at each network layer, which encode learned features of the syndrome history---and recompute only the temporal boundary region affected by each new round, whose extent depends on the number of convolutional blocks. A running sum preserves temporal averaging over the complete available history. The RTT processes each new round by reusing stored representations of earlier rounds and updating running sums used to average the token features for the final prediction. Moreover, we implemented the streaming version of Frontier in WebAssembly (WASM), which is supported by our real-time control stack~\cite{moses2023race, ryan2022implementing, brown2022quantum}  (see Methods). 
 Fig.~\ref{fig:decoding_stats} shows the processing time per round for Frontier and the neural network decoders. Frontier can keep up with syndrome generation, and the neural decoders offer low-latency inference compatible with real-time operation on the trapped-ion device. Ultimately, a hybrid approach like that proposed in Ref.~\cite{olle2026fast} could prove to be an effective real-time decoding strategy, where a neural architecture resolves easy syndromes and an algorithmic decoder resolves what remains.

\noindent\textbf{Distance Scaling Experimental results:} To find the logical error rate per syndrome extraction round (LER), one can run memory experiments for different number of rounds $r$, estimate the failure probability $P_L(r)$ as a function of $r$ and extract the LER from fitting $P_L(r)$ to the experimental data. However, for extremely small values of LER where failures are significantly rare, estimating $P_L(r)$ with a reasonable error bar even for a single value of $r$ is already challenging, making it impractical to estimate $P_L(r)$ for multiple values of $r$ with error bars that are good enough for fitting. Here we estimate LER from a single $r=20$ point instead, 
using the approximate formula ${\rm LER}=\frac{n_F/n}{r+1/2}$, with $n_F$ and $n$ denoting the total number of failures and total number of shots (summed over all bases) respectively, and $r=20$ denoting the number of syndrome extraction cycles. The factor of $1/2$ accounts for the logical state preparation error, as it is essentially half of a syndrome extraction cycle (see Supplementary Material). The detailed experimental results are listed in Table~\ref{tab:shot_stats}. All decoders show decreasing logical error rates with increasing code distance. The algorithmic decoders are technically indistinguishable within 1-$\sigma$ error bars, and the two neural-network decoders achieve comparable logical performance to algorithmic decoders. By fitting the experimental data to the ansatz $\log(\text{LER}(d))=-\frac{\log(\Lambda)}{2} d + c$, we find suppression factors of $\Lambda_{\text{Tesseract}}=12^{+3}_{-2}$, $\Lambda_{\text{Frontier}}=11^{+2}_{-2}$, $\Lambda_{\text{CNN}}=7^{+1}_{-1}$ and $\Lambda_{\text{RTT}}=8^{+1}_{-1}$ using the Tesseract, Frontier, CNN and RTT decoders respectively.
Figure~\ref{fig:main}(c) shows the Tesseract decoding results on hardware and simulation data. 

\begin{figure}[t]
    \centering
    \includegraphics[width=\columnwidth]{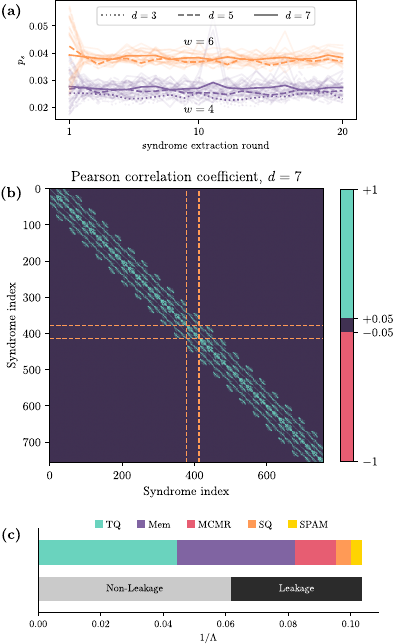}
    \caption{{\bf Syndrome error analysis} a) Average non-trivial syndrome 
    density ($p_s$) per round. Faded lines correspond to specific stabilizers, and bold lines indicate shot-averaged quantities for each distance. Syndromes corresponding to weight $w=4$ and weight $w=6$ stabilizers are averaged separately. b) Correlation between detector firings. The dashed lines show the range of detectors corresponding to the $10$th round of syndrome extraction, showing that correlations with the $10$th round are only significant in the $9$th and $11$th rounds. A discrete color map is used for better contrast (see Supplementary Material for analogous plots with continuous color maps for all distances). c) Error budget for $1/\Lambda$, estimated from simulation.
    }
    \label{fig:syndrome_analysis}
\end{figure}

\noindent\textbf{Error analysis:}
Beyond the logical error rate, the memory experiments produce a trove of syndrome data that provides experimental data  about the underlying hardware error sources and their correlations.
Fig.~\ref{fig:syndrome_analysis}(a) shows the probability of measuring a non-trivial syndrome $p_s$ at each round for different distances. Syndromes corresponding to weight $4$ and $6$ stabilizers are separated as they naturally have different probabilities of returning $1$. The light lines correspond to individual stabilizers and the dark lines correspond to their average at each round. As is clear from Fig.~\ref{fig:syndrome_analysis}(a), average $p_s$ remains stable in the span of a shot as well as between different distances. 
Early in the data taking, selected stabilizers were elevated at certain rounds.
These syndrome signatures exposed inaccuracies in the qubit phase tracking on Helios,  which were remedied with software improvements removing the spurious syndromes (see Supplementary Material). We include data both with and without the phase tracking improvement in the main data sets, as the additional phase errors were routinely corrected in QEC and did not significantly affect the LER. However, this example illustrates how syndrome ``error spectroscopy" can provide powerful insights into hardware error mechanisms that may evade traditional diagnostics and enable the improvement of quantum processors.

Fig.~\ref{fig:syndrome_analysis}(b) shows the Pearson correlation $\text{Cov}(X,Y)/\sqrt{\text{Var}(X)\text{Var}(Y)}$
between different syndromes in the $d=7$ experiments.  The dashed lines show the range of syndromes that belong to the $10$th syndrome extraction round. The data demonstrates that each syndrome is only correlated with the syndromes belonging to the same round and the neighboring rounds, indicating the absence of significant time-correlated errors.  

Lastly, we use the Helios-1 QPU emulator to estimate the contributions of different error sources to $1/\Lambda$, following the procedure in Ref.~\cite{google2021exponential}. As expected, two qubit gate error is the dominant noise source at $43\%$, followed by $36\%$ memory errors mostly related to transport and $13\%$ from mid-circuit measurement and reset cross talk (MCMR). Leakage is a share of each of these noise channels. If we add the leakage contribution to all channels, it amounts to $41\%$ of $1/\Lambda$, which is mostly turned into depolarizing noise in the qubit space  using leakage repump. 

\noindent{\bf Clifford Randomized Benchmarking:}
\begin{figure}
    \centering
    \includegraphics[width=\columnwidth]{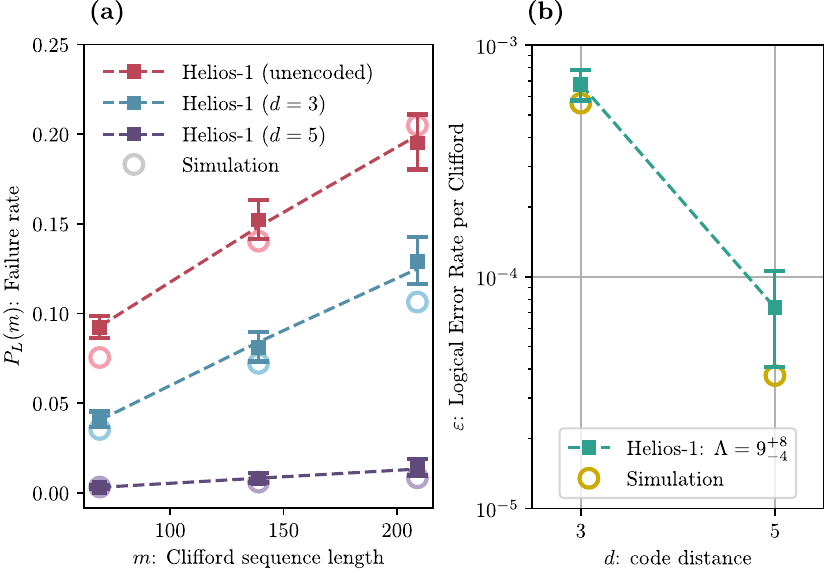}
    \caption{{\bf Clifford 2QRB} a) the Failure rate $P_L(m)$ of unencoded and encoded Clifford 2QRB circuits for different 2QRB sequence lengths $m$. b) Logical error rate per Clifford gate $\varepsilon$ for distances $d=3,5$. We extract $\varepsilon$ assuming the decay form $P_L=A(1-\frac{4}{3}\varepsilon)^m+\frac{3}{4}$. The error bars on simulation data points are smaller than their marker and hence are omitted.}
    \label{fig:2qrb}
\end{figure}
We study the performance of logical Clifford two-qubit randomized benchmarking~\cite{mayer2024, berthusen2026} (2QRB) circuits when each qubit is encoded in a color code block of distances $d=3,5$. We also run the 2QRB circuits unencoded for comparison. A 2QRB circuit of length $m$ consists of a sequence $C_1, C_2, \cdots, C_m$ of two-qubit Clifford gates, each chosen uniformly at random, applied to the $\ket{00}$ state and followed by the Clifford gate $C_\text{inv}=(C_m\, C_{m-1}\cdots C_1)^{-1}$ and then measurement in the computational basis. We compile each $C_i$ in terms of $H$, $S$, $S^\dagger$, $\text{SWAP}$ and $\text{CNOT}$ gates plus the Paulis. The $\text{SWAP}$ gate is implemented in software through relabeling in both encoded and unencoded circuits. For the encoded circuits, since logical gates are implemented purely transversally, there is no need to extract syndromes after each logical gate to prevent the spread of errors within each code block. Instead, we add SWAP-superdense syndrome extraction circuits after each $7$ logical Clifford gates, as the syndrome extraction circuit has itself CNOT-depth of $7$. Lastly, we run $37$, $3$ and $1$ 2QRB circuits in parallel in each shot for unencoded, $d=3$ and $d=5$ experiments respectively, which makes all experiments utilize roughly the same number of physical qubits in each shot, similar to the memory experiments. $d=7$ 2QRB circuits require more qubits than what is available on Helios-1 and hence are not considered in this section. 

Fig.~\ref{fig:2qrb}(a) shows the experimentally observed failure rates $P_L$ when sampling $2220, 1110$ and $666$ circuits of sequence lengths $m=69, 139$ and $ 209$ respectively. Each sampled circuit has been run once unencoded, and once encoded in $d=3,5$ color codes. The encoded circuits for $m=69, 139$ and $209$ have $9$, $19$ and $29$ rounds of syndrome extractions respectively, and have been decoded offline using Tesseract decoder for optimal performance. By fitting the experimental data to the ansatz $\log(\frac{3}{4}-P_L(m))=m\,\log(1-\frac{4}{3}\varepsilon)+c$ (see Supplementary Material), we find error rates per Clifford gate of $\varepsilon_\text{unenc.}=9.5^{+0.6}_{-1.4}\times 10^{-4}$, $\varepsilon_3=7^{+1}_{-1}\times 10^{-4}$ and $\varepsilon_5=7^{+3}_{-3}\times 10^{-5}$ for unencoded, $d=3$ and $d=5$ circuits respectively. The observed logical error rates correspond to the suppression factor of $\Lambda_{3/5}^{\text{Logic}}=9^{+8}_{-4}$ (see Fig.~\ref{fig:2qrb}b). 

\noindent{\bf Discussion:}
Assuming a Clifford plus T model of universal quantum computation with T gates teleported via magic state injection, a general circuit encoded in color code will have three non-trivial components: magic state preparation, two qubit Clifford gates and syndrome extraction gadgets. This work demonstrates high accuracy, below threshold implementation of the last two components on an ion-trap device. What remains is demonstrating below threshold scaling of high-fidelity color code magic state preparation~\cite{bravyi2005universal, souza2011experimental, brown2023advances, sales2025experimental, chamberland2019fault, chamberland2020very, gidney2024magic, hetenyi2026constant, rosenfeld2025magic,daguerre2025code, daguerre2025experimental}. Furthermore, teleporting T gate via magic state injection requires real-time decoding, which the results in this work suggests is already viable for near optimal decoders. Hence a natural next step is to demonstrate below threshold behavior on a full Clifford plus T circuit encoded in color code using real-time decoding of the syndromes.

While this work highlights the improvement leakage repump has on logical fidelity, it is interesting to study how much of this improvement can be reclaimed by simply using a better decoder which can reason about leakage errors~\cite{bultink2020protecting, mude2025accurate, nishio2026qubit, computing2026quantum}. Given that leakage repump is not perfect and some fraction of leaked qubits remain leaked, such a decoder can prove useful even while using leakage repump. It is worth noting that this will also help the decoding latency of the algorithmic decoders, since a simple DEM which does not include leakage makes the decoder search for higher order errors to solve for a syndrome related to leakage. 

Both neural network decoders have been trained on emulator data. Although a great effort has been put into arriving at a faithful emulation, there could still be error channels on the device that are not captured by the emulation. Furthermore, the emulator data has been generated using stabilizer simulations which only approximates the effects of non-stabilizer errors like coherent memory errors on the device. As such, it is interesting to study fine tuning ML decoders on hardware data to improve their accuracy. Since it is challenging to generate very large amount of hardware data on an ion-trap device for training, one needs to find an efficient fine-tuning method that works well on small data sets.

Logical fidelity estimation via direct sampling will become harder as the hardwares improve and scale. Therefore it is inevitable to use alternative methods for logical benchmarking which have better scaling in terms of required sample size~\cite{xiao2026situ, zheng2026efficient}. It is interesting to study these methods in the context of this experiment and compare their estimate with what direct sampling yields. Another closely related direction is to use syndrome data to learn the circuit error models~\cite{blume2025estimating} which in turn can improve both decoding and logical benchmarking accuracy. This direction could prove fruitful for studying error-floors on Helios as well.

\section*{Acknowledgments}
We acknowledge the entire Quantinuum team for innumerable contributions towards the design and operation
of the Helios quantum computer. We specially thank David Amaro and Pablo Andres-Martinez for helpful discussion and feedback on the manuscript.

\section*{Methods}

\subsection{Leakage repump}
\label{sec:leakage_repump}

The goal of hardware-level leakage reduction is to repump leaked population into the computational subspace, thereby converting leakage into errors that can be handled by standard qubit-based quantum error correction techniques. For Helios QPUs, each qubit is encoded in a pair of first-order magnetically insensitive levels in the ${}^2S_{1/2}$ electronic ground state of a $^{137}\mathrm{Ba}^{+}$ ion, namely the `clock' states $\{\lvert F{=}1, m_{F}{=}0\rangle,\lvert F{=}2, m_{F}{=}0\rangle\}=\{|0\rangle,|1\rangle\}$, while the remaining six hyperfine levels form the leakage subspace~\cite{ransford2026}.

To address all six leakage levels, we employ an optical pumping scheme that extends the state preparation technique presented in Ref.~\cite{an2022}. Each cycle of the repump scheme, depicted in Fig.~\ref{fig:repump_overview}(a), consists of eight state-selective $1762\,\mathrm{nm}$ pulses (one for each $F=2$ leakage level and two for each $F=1$ leakage level) that sequentially drive electric-quadrupole transitions between the leakage subspace and the three $F=1$ levels in ${}^2D_{5/2}$. Selection rules and the hyperfine splitting of ${}^2D_{5/2}$ allow individual addressing of leakage levels, leaving the qubit clock states largely unperturbed, similar to the approach of Ref.~\cite{hayes2020}.  To further minimize off-resonant coupling to the qubit levels, the amplitude of each $1762\,\mathrm{nm}$ pulse is shaped to suppress spectral sidebands that could erroneously drive transitions involving the computational subspace.

Following each $1762\,\mathrm{nm}$ pulse, $614\,\mathrm{nm}$ light
drives ${}^2D_{5/2}\leftrightarrow {{}^2P_{3/2}}$ via an electric-dipole transition, from which population spontaneously decays to ${}^2S_{1/2}$, ${}^2D_{5/2}$, and ${}^2D_{3/2}$ with branching ratios $0.74$, $0.23$, and $0.03$, respectively~\cite{Zhang2020}. Population that returns to ${}^2D_{5/2}$ is re-excited by the same $614\,\mathrm{nm}$ light, while population in ${}^2D_{3/2}$ is cleared by $650\,\mathrm{nm}$ light, which is left on throughout the repump sequence.

Figure~\ref{fig:repump_overview}(b) shows the leaked population as a function of the number of repump cycles, which we use to characterize leakage suppression. To prepare the leaked population, we use Doppler cooling to leave each ion in an incoherent mixture of the eight ${}^2S_{1/2}$ hyperfine levels, so that most of the population starts in the leakage subspace. We then apply $n$ repump cycles, followed by a ternary measurement that classifies each shot as $|0\rangle$, $|1\rangle$, or leaked~\cite{ransford2026}.

We extract the leakage suppression by independently fitting the leaked fraction of the 16 qubits in the eight gate zones. The leaked fraction $P_i(n)$ of qubit $i$ after $n$ repump cycles is estimated from $1{,}000$ shots and fit to $P_i(n) = a_i r_i^n + b_i$ by nonlinear least squares with all parameters free, weighted by binomial uncertainties. Here $r_i$ is the decay factor, $b_i$ is the leakage floor, and $a_i$ is the initial leaked fraction above the floor. Averaging over all qubits gives $a = 0.786(3)$, $b = 4.9(6)\times10^{-3}$, and $r = 0.529(3)$, corresponding to a suppression of $1/r = 1.89(1)\times$ per cycle with uncertainties representing one standard deviation. At the three-cycle operating point, averaging the per-qubit fitted ratios $P_i(0)/P_i(3)$ gives a suppression of $6.5(1)\times$, corresponding to a relative residual leaked fraction of $0.153(2)$.  These results are in rough agreement with numerical simulations of the repump sequence, which predict that after three cycles a qubit starting in a uniform mixture over the leakage subspace $L$ ends in $|0\rangle$, $|1\rangle$, or $L$ with probabilities $p(0|L)=0.78$, $p(1|L)=0.10$, and $p(L|L)=0.12$. Notably, simulations predict that the majority of repumped population returns to~$|0\rangle$.

\subsection{Decoding methodology}
\subsubsection{Frontier Streaming Decoder}
The original formulation of Ref.~\cite{leverrier2026} is naturally an offline decoder: the complete measured syndrome $s$ is supplied to the decoder before the ordered recursion begins. The deadline ordering used there starts from the natural ordering of the fault locations and rearranges variables so that detector rows tend to close early while keeping the active boundary small. Importantly for real-time decoding, this ordering remains close to the temporal circuit ordering; the original work explicitly notes this as desirable for future online implementations. This motivates the streaming formulation used here, in which the same Frontier recursion and ordering are retained but detector outcomes are supplied incrementally as the experiment proceeds.

We consider streaming decoders parameterized by a temporal window $W$. Here, $W$ denotes the number of syndrome rounds available when a round at the leading edge of the window is processed. In \textit{zero-lag} streaming, corresponding to $W=1$, detector round $R$ is processed as soon as it is observed. No future detector outcomes are available when the current round is processed, although information from all previous rounds remains represented in the persistent Frontier state. In \textit{fixed-lag} streaming, with $W > 1$, Frontier processing trails syndrome acquisition by $L = W-1$ rounds. The part of the model associated with round $t$ is processed only after syndrome data through round $t + W-1$ have been acquired. The additional lookahead changes the information available when the practical finite-$K$, 
$\Delta$ decoder performs its pruning. In practice, this allows later detector outcomes to eliminate incompatible hypotheses before they occupy the retained frontier, leading to tighter decoding times, see below. 

The streaming construction used here and below in the neural network decoders should also be distinguished from the standard $(w,c)$ overlapping-window decoder~\cite{dennis2001topological, Skoric2023, huang2024}. In that formulation, $w$ syndrome rounds are decoded together, but only the correction associated with the first $c \le w$ rounds is committed; the remaining tentative correction is discarded, the committed correction is propagated into the subsequent syndrome, and the decoding window is advanced by $c$ rounds before the procedure is repeated. Our fixed-lag Frontier decoder has a superficially similar information horizon; however, it has different semantics. Frontier does not independently decode each $W$-round window or commit a physical correction whenever the window advances. Instead, the retained distribution over Frontier states persists across the entire circuit, so information accumulated before the current window is carried forward in the soft decoder state. In this sense, $W$ in our streaming decoder specifies a lookahead or processing lag, rather than the size of a repeatedly decoded-and-committed temporal block. The streaming constructions here could be turned into a conventional $(w,c)$ windowed decoder by forcing a hard commit of the first $c$ rounds after each $w$-round lookahead interval, updating the residual syndrome accordingly, and discarding posterior information that is no longer compatible with the committed region.

We implemented the streaming version of Frontier in Rust and optimized it for low-latency execution in WebAssembly (WASM)~\cite{moses2023race, ryan2022implementing, brown2022quantum}. 
The Quantinuum’s real-time software stack has support for calling arbitrary WASM functions from Guppy and thus supports this implementation, though the development of this implementation was done after the experiments.
To that end, syndrome data derived from previously acquired Helios-1 quantum measurements were hardcoded into Guppy programs and supplied to the decoder one round at a time. 
We measured individual WASM execution times for streaming syndrome processing and the final correction call across all 66,440 records at code distances.
These measurements characterize decoder execution on a local laptop rather than end-to-end feedback latency; integration into the real-time control stack introduces additional runtime and communication overheads. Nevertheless, the measured latencies demonstrate the potential for real-time decoding. 

\subsubsection{Neural Network Decoders}
We train a 3D CNN and a round-token transformer (RTT) on the same datasets generated by the Helios-1 emulator,  using the same circuits that were run on Helios-1, but with the noise intensity increased by a factor of four to improve error statistics during training. Both predict whether the measured logical observable has flipped and are evaluated on hardware data without fine-tuning. They provide complementary ways to combine spatial and temporal syndrome information, with comparable logical performance summarized in Table~\ref{tab:shot_stats}. Both neural decoders are able to operate in real-time on Helios QPUs, with a latency that is much lower than the relevant operations timescale. The CNN and RTT process the syndromes for each round sequentially, with average per-round GPU compute times of $t_{\mathrm{round}} \approx 0.26, 0.36, 0.52$ ms for the CNN and $t_{\mathrm{round}} \approx 0.22, 0.25, 0.35$ ms for the RTT at code distances $d=3,5,7$, respectively, on the NVIDIA H100 80GB GPU (see Fig.~\ref{fig:decoding_stats}).  The following provides a general overview of each decoder while more details on their architecture and training are provided in the Supplementary Material.

\textit{3D CNN:}
The CNN embeds detector events and detector-type labels on a tensor indexed by syndrome-extraction round and two integer lattice coordinates (Fig.~\ref{fig:2d_array_mapping}). This construction applies generally to topological codes whose check locations form a Euclidean lattice, with  periodic or open boundary conditions: a reciprocal basis extracts the integer coordinates, with distinct checks within a unit cell assigned separate channels. For a lattice with $N$ spatial dimensions, stacking syndrome-extraction rounds yields an $(N+1)$-dimensional array for processing by an $(N+1)$-dimensional CNN. Here $N=2$: the color-code check locations lie on the triangular lattice dual to the honeycomb tiling. The core convolutional architecture follows a ResNet-style bottleneck design, using symmetrically padded three-dimensional convolutions to learn local space--time correlations~\cite{he2016deep,hara2018spatiotemporal}. The readout averages features over time, selects a spatial readout slice after a shared local  multi-layer perceptron (MLP), and averages over the slice. A final MLP maps the pooled representation to a logical-flip logit. The logit provides an unnormalized score that the logical observable of the initial quantum state has flipped. 

For streaming inference, we cache intermediate activations and recompute only the temporal boundary region affected by each new syndrome round, whose extent grows with the number of convolutional blocks. A running sum preserves temporal averaging over the complete available history, with zero padding beyond the currently available history. New measurements replace this padding and can therefore change earlier feature representations near the temporal boundary. This supports online predictions with a cost that depends on the number of convolutional blocks.

\textit{Round-token transformer (RTT):}
The inputs provided to the RTT network are the syndromes extracted at each round.
Inside the network, the syndromes are grouped into a vector and processed by a MLP encoder (or optionally a graph). They are subsequently embedded together with a learned start vector, which encodes the temporal distance of the round from the beginning of the experiment and saturates after the first few rounds. This feature allows the model to better generalize and to be more easily trained on multiple set of experiments with different number of rounds. The result of this encoding is one token per syndrome extraction round. The tokens are then passed through a stack of causal transformer blocks: each of them implements multi-head causal self-attention, where only the present round token is attended. Moreover, in order to include memory of the past, the keys and values of the tokens are cached in memory and reused. A learned time bias is also added before projecting the tokens and processing them with a small feed-forward network, wrapping both processes with residual connections. \\
After the final transformer block and a layer normalization, a gating layer is applied to the tokens, which are weighted by a gate which determines how each one is informative.  
Each token is then added into a running accumulator, so that the network carries a fixed-size memory of the whole history regardless of how many rounds have elapsed. A summary vector, resulting from a gated average concatenated with the plain average of the tokens, is created and passed to the output head, a small MLP which returns a single logit.

\bibliographystyle{apsrev4-2}
\bibliography{refs}


\clearpage
\newpage

\appendix
\onecolumngrid
\begin{center}
  \Large\bfseries Supplementary Material 
\end{center}

\section{Leakage repump}
\subsection{Concept of operations}
\label{sec:repump_conops}

\begin{figure}[h]
    \centering
    \includegraphics[width=1.0\columnwidth]{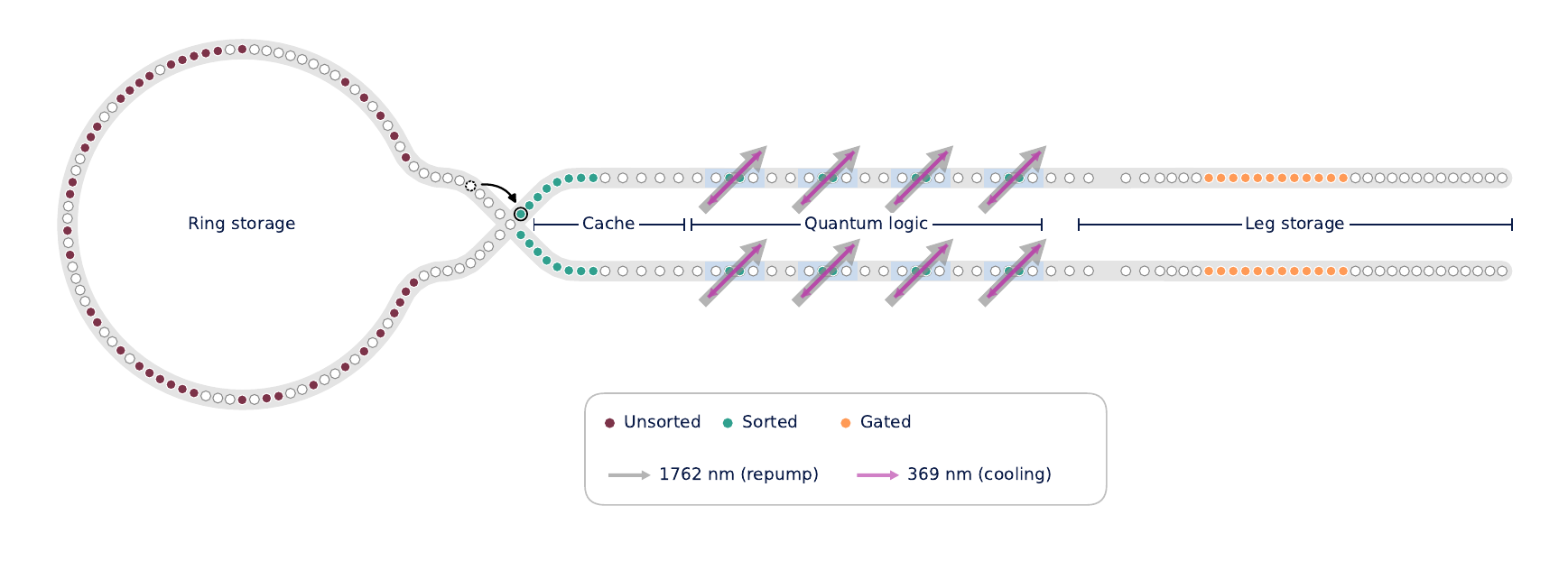}
    \caption{Diagram of the Helios ion trap showing how the leakage repump scheme is implemented in a circuit. Qubits are colored by their status: unsorted (ring), sorted (cache), or gated (leg storage). Gray arrows denote 1762~nm leakage-repump light and pink arrows the counter-propagating 369~nm ground-state cooling light, both addressing the two qubits in each zone center in the quantum logic region. Shaded blue regions mark the eight gate zones in the quantum logic region. The black arrow indicates transporting a qubit from the ring into an empty well in the cache. Not shown in this diagram are the 614~nm and 650~nm sheet beams that illuminate the entire trap.}
    \label{fig:trap}
\end{figure}

The concept of operations for Helios was first described in Ref.~\cite{ransford2026}. Since then, we have upgraded Helios with mid-circuit leakage repump, which users can enable in their Guppy programs. We implement this feature without increasing circuit runtime by parallelizing leakage repump with the two most time-consuming operations on Helios: ground-state cooling and ion transport. Ground-state cooling and leakage repump light are applied simultaneously to a batch of qubits immediately before the batch is gated in the quantum logic region. One cycle of leakage repump takes ${\sim}340~\mu$s, so the three-cycle operating point (see Methods) fits within the ground-state cooling time, which is roughly twice as long. Cooling and repump run while qubits are sorted from the ring for the next batch. In this way, cooling, repump, and transport can all proceed in parallel.

Figure~\ref{fig:trap} illustrates this parallelization partway through an example circuit: one qubit is transported out of the ring into the cache while the qubits already in the quantum logic region are simultaneously cooled and repumped, and a previously gated batch is moved to the leg storage. Once cooling finishes, the repumped qubits are gated and flushed to the leg storage, and the batch staged in the cache shifts in to take their place.

\subsection{Benchmarking}
\label{sec:repump_benchmarking}

\begin{figure}[t]
    \centering
    \includegraphics[width=\columnwidth]{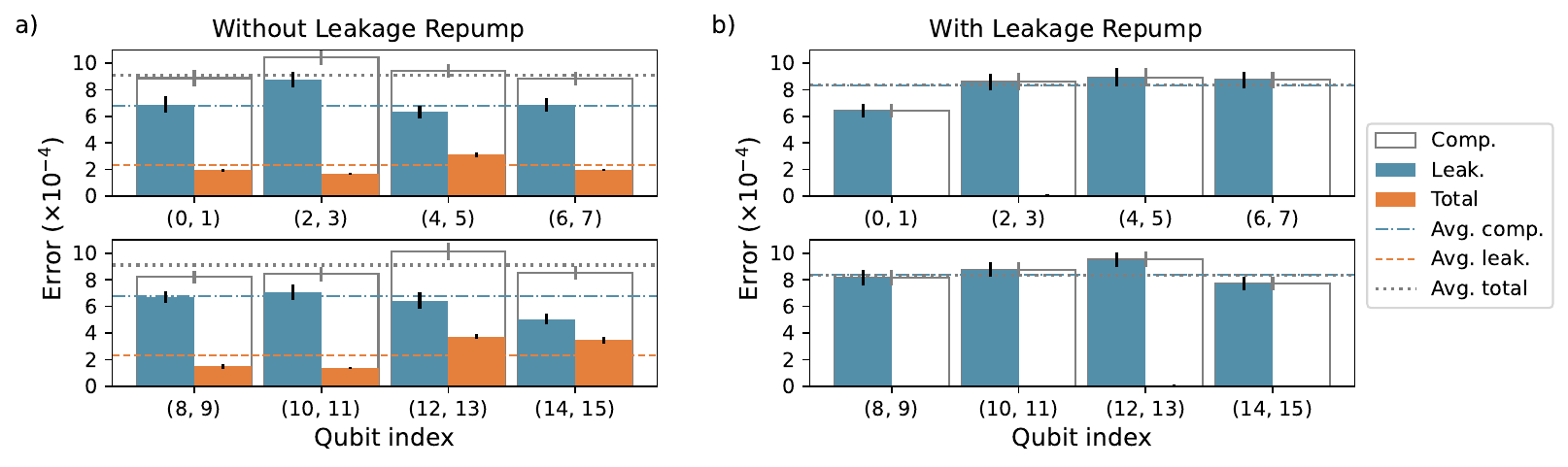}
    \caption{Two-qubit randomized benchmarking (a) without and (b) with leakage repump.  The two-qubit average fidelity is decomposed into leakage and computational errors for 8 qubit pairs in the eight gate zones, with qubit 0-7 in the top and 8-15 in the bottom zones, respectively, ordered from left to right.  }
    \label{fig:2q_benchmarking}
\end{figure}

We quantify errors induced by the leakage repump operation on the computational subspace by performing two-qubit randomized benchmarking (2QRB) without and with leakage repump as shown in Fig.~\ref{fig:2q_benchmarking}(a) and (b), respectively.  Circuits of lengths $l\in\{10, 100, 200, 400\}$ were run with 1,000 shots total per length.  Applying the ternary measurement operation \cite{ransford2026} at the end of each circuit enables the separation of leakage and computational errors, following the methods of \cite{Chen2025}.  From the leakage population versus sequence length, we extract the leakage rate per Clifford $\tau$.  The remaining population is fit to the decay model $p(l) = Ar^l+1/4$ where $A$ is the state preparation and measurement (SPAM) parameter, $r = 1 - \lambda - \tau$ is the depolarizing parameter, and $\lambda$ is the computational error.  Using the fact that there are
1.5 two-qubit gates per Clifford on average \cite{ransford2026}, the average infidelity per two-qubit gate (including leakage) is then computed as $\epsilon_{\mathrm{avg, 2Q}} = \lambda/2 + \tau_\mathrm{2Q}$, where $\tau_\mathrm{2Q} = 2\tau/3$.  As shown in Fig.~\ref{fig:2q_benchmarking}, we find that without leakage repump $\epsilon_{\mathrm{avg, 2Q}}=9.1(2)\times10^{-4}$ and $\tau_{\mathrm{2Q}}=2.3(1)\times10^{-4}$, where leakage errors arise due to spontaneous emission and memory errors \cite{ransford2026}.  With leakage repump, we find $\epsilon_{\mathrm{avg, 2Q}}|_\mathrm{w/LR}=8.4(2)\times10^{-4}$ and statistically insignificant leakage errors.  

The leakage repump operation therefore leads to a net {\it improvement} of the two-qubit gate infidelity by $7(2)\times10^{-5}$, which excludes application of methods such as interleaved randomized benchmarking (e.g., as done in Ref.~\cite{hayes2020}) to isolate the error induced by the leakage repump operation.  Instead, we introduce a physical argument to isolate this error, denoted $\epsilon_q$, from the overall change in fidelity.  Consider a leakage event ocurring between Cliffords during a 2QRB experiment.  The immediate repump converts this event into a Pauli error, which is twirled by the subsequent Cliffords into an effective depolarizing channel.  The non-leaked qubit in the pair will also be depolarized by virtue of the random Cliffords.  Consequently, the repumped leakage event is equivalent to a Pauli error, contributing decay to the survival probability at a rate $3\tau/4$.  Since this is a computational error, it is equivalent to an effective depolarizing parameter via the replacement $\lambda\to\lambda + 4\tau/5$.  Subtracting infidelities with and without leakage repump, we find $\epsilon_{\mathrm{avg, 2Q}} - \epsilon_{\mathrm{avg, 2Q}}|_\mathrm{w/LR} = (2/5)\tau_{2Q} - \epsilon_q$, from which we estimate an improvement of the infidelity due to leakage conversion of $9(1)\times10^{-5}$ countered by error induced on the computational subspace of $\epsilon_q= 2(2)\times10^{-5}$, which serves as an upper bound. 

Previous work identified polarization impurity, beam misalignment, and off-resonant transitions as sources of qubit error during leakage repump~\cite{hayes2020}. With pulse shaping to suppress off-resonant transitions, we expect these errors to be suppressed below $10^{-7}$ per cycle. The remaining error is expected to be dominated by phase and intensity noise on the $1762$~nm light. Phase noise can off-resonantly drive qubit population to ${}^2D_{5/2}$, causing leakage, while intensity noise causes fluctuating AC Stark shifts.

\section{Bell Flag and Superdense Syndrome Extraction Circuits}
\begin{figure}[h]
    \centering
    \includegraphics[width=0.7\textwidth]{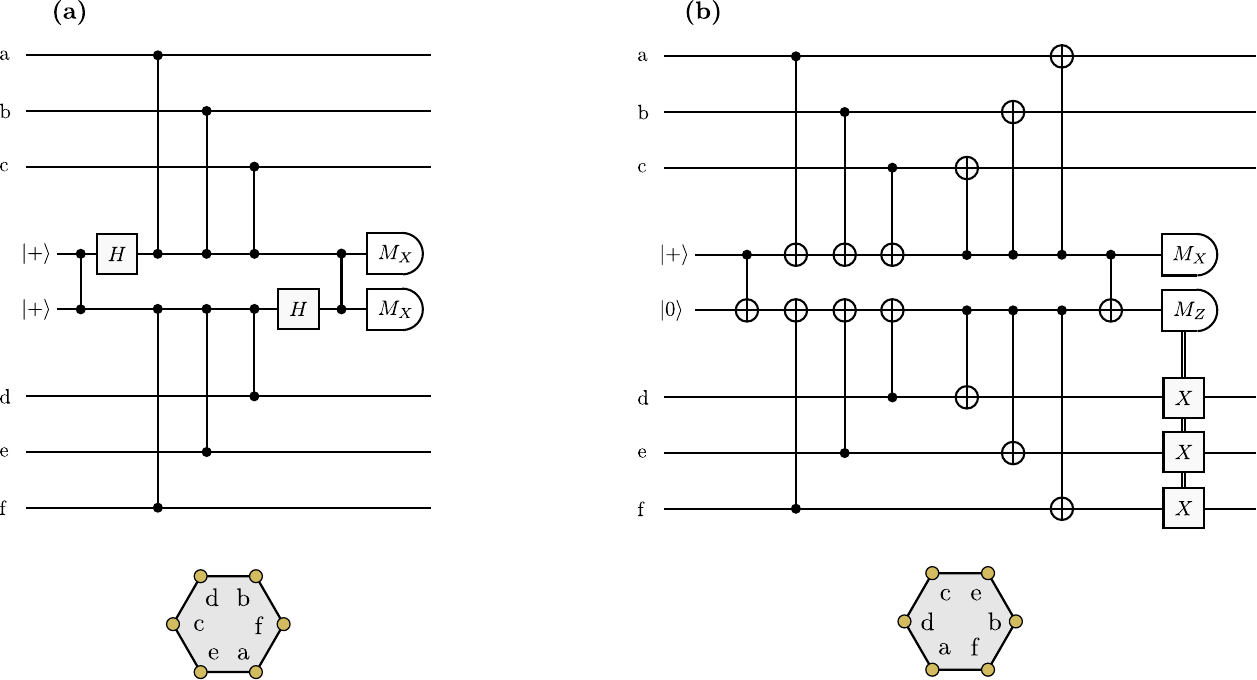}
    \caption{{\bf Bell-flag and superdense syndrome extraction circuits} a) Bell-flag syndrome extraction circuit. The top measurement corresponds to the product of $Z$ on data qubit wires, and the bottom measurement is a flag measurement. b) Superdense syndrome extraction circuit. The bottom figures show the mapping between wires in the circuits and the data qubits on each face, for each circuit. }
    \label{fig:bell_and_superdense}
\end{figure}

The Bell-flag syndrome extraction circuit~\cite{baireuther2019neural} shown in Fig.~\ref{fig:bell_and_superdense}(a) is used for state preparation. The circuit shown in Fig.~\ref{fig:bell_and_superdense}(a) is for measuring $Z$ stabilizers. The circuit for measuring $X$ stabilizers is obtained by adding Hadamard gates on data qubit wires at the beginning and end of this circuit. For four qubit faces, we simply drop the missing wires and their associated gates from the circuit. We use this circuit for state-preparation, since only half the stabilizers need to be measured at that stage, and using the Bell-flag circuit results in a shallower state preparation circuit. The syndrome qubit measurement outcomes define the Pauli frame of the code state, and the flag qubit measurement outcomes are fed to to the decoder when decoding the entire circuit.

Fig.~\ref{fig:bell_and_superdense}(b) shows the superdense syndrome extraction circuit~\cite{gidney2023new} and the mapping between circuit wires and qubits on each face. The SWAP-superdense circuit shown in Fig.~\ref{fig:main}(b) is obtained from Fig.~\ref{fig:bell_and_superdense}(b) by using the circuit identity $\text{CNOT}_{1,2}\, \text{CNOT}_{2,1}=\text{SWAP}\, \text{CNOT}_{1,2}$ on the two pairs of middle CNOTs connected to the two syndrome qubits. 

\section{Syndrome Density Per Round, Before and After Phase Tracking Improvement}
\begin{figure}[h!]
    \centering
    \includegraphics[width=\textwidth]{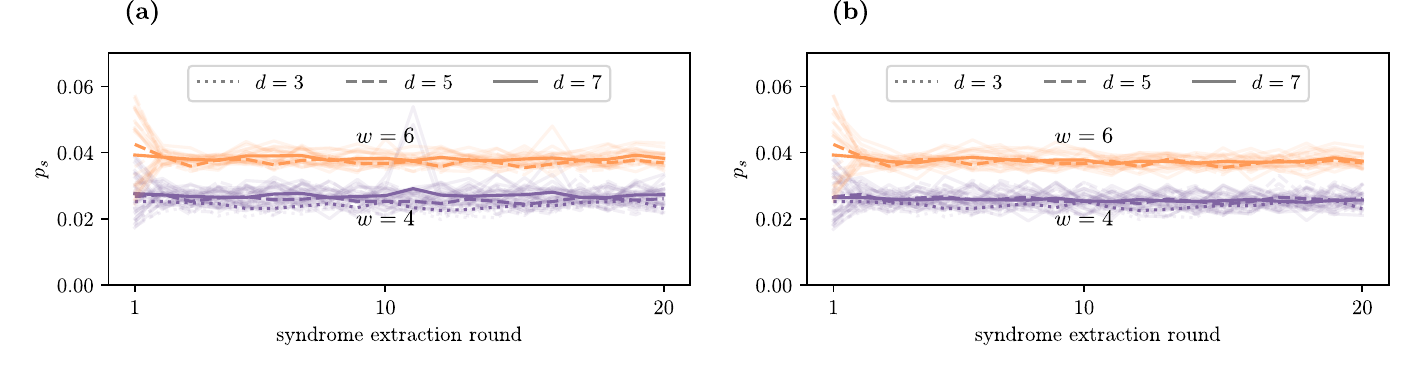}
    \caption{{\bf Average Syndrome density per round} a) including all the shots, b) including only the shots taken after phase tracking improvements }
    \label{fig:all_vs_new_p_d_vs_time}
\end{figure}

Fig.\ref{fig:all_vs_new_p_d_vs_time}a and Fig.\ref{fig:all_vs_new_p_d_vs_time}b shows the average syndrome density per round for all shots and shots that were taken after resolving an edge case in the software tracking of a qubit's phase accumulated during transport. As is clear, the software change made the anomalous elevated syndromes vanish. 

\section{Correlations between different syndromes}
\begin{figure}[h]
    \centering
    \includegraphics[width=\textwidth]{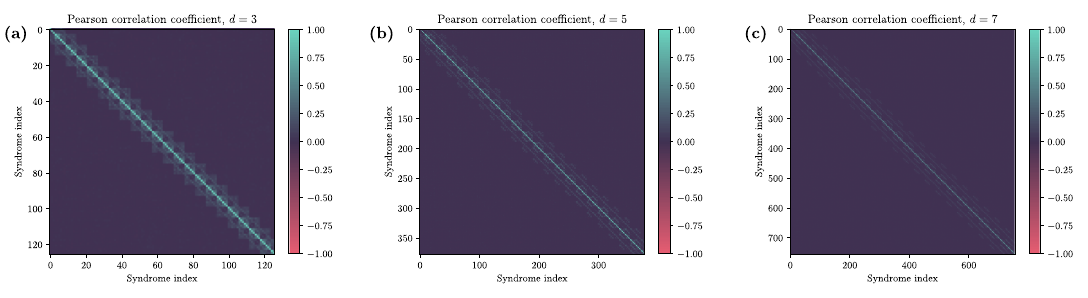}
    \caption{{\bf Pearson Correlation Coefficient} The correlation between all detectors in the memory experiment for a) $d=3$, b) $d=5$ and c) $d=7$ }
    \label{fig:corr_soft_all}
\end{figure}

Fig.~\ref{fig:corr_soft_all} shows the Pearson correlation coefficient between syndromes in $d=3,5,7$ memory circuits. Larger circuits have more syndromes, hence the horizontal and vertical ranges grow with distance. However, for all distances, the non-trivial correlations group into neighboring blocks, which correspond to neighboring rounds of syndrome extraction. Fig.~\ref{fig:syndrome_analysis}(b) is the same as Fig.~\ref{fig:corr_soft_all}(c), but with hard cut-offs to enhance the contrast of non-trivial correlations. 
\section{Logical 2QRB Decay form}
Let $P_m$ denote the probability of failure for a length $m$ 2QRB circuit and let $\varepsilon$ denote the failure probability of a two-qubit gate. Furthermore, assume a two-qubit gate failure will flip first, second or both bits of the final measurement with equal probabilities. By conditioning on the first two-qubit gate failure in a length $m$ 2QRB experiment, we arrive at the following recursive relation:
\begin{align}
    1-P_m=(1-\varepsilon)(1-P_{m-1})+\varepsilon\frac{P_{m-1}}{3},
\end{align}
which can be solved to arrive at the $P_m=(P_0-\frac{3}{4})(1-\frac{4}{3}\varepsilon)^m+\frac{3}{4}$ decay form.
\section{Neural-Network Decoders}

We train two complementary neural-network decoder models on emulator data. The first is a three-dimensional convolutional neural network (3D CNN). Earlier work used fully convolutional 3D networks as local pre-decoders for surface-code circuit noise~\cite{chamberland2023local3d} and demonstrated fully convolutional decoders that generalize to large space--time volumes~\cite{gicev2026fully3d}. More recently, a CNN attained near-optimal accuracy on surface and quantum-LDPC codes while retaining accelerator-friendly throughput~\cite{gu2026cascade}. This progression motivates the CNN as a strong, spatially structured decoder for our color code memory experiments. Our contributions are a general syndrome embedding method and a streaming implementation of the 3D CNN. The embedding maps check locations on Euclidean lattices to arrays using a reciprocal lattice basis, with separate channels for distinct checks within each unit cell and learned embeddings of detector outcomes and types. This construction preserves local spatial structure and extends naturally to higher-dimensional lattices. For streaming inference, we cache intermediate activations, update only the temporal boundary region affected by each new syndrome round, and maintain a running sum for temporal averaging. This allows online predictions incorporating the complete available syndrome history without recomputing the full space-time volume at each round.

The second model is a streaming round-token transformer. Transformer-based QEC decoders predate our model: Ref.~\cite{choukroun2024deepqec} adapted a parity-check-aware error-correction-code transformer to stabilizer codes, and Ref.~\cite{wang2023transformerqec} used global self-attention over a position-encoded space--time syndrome and transferred models between surface-code distances. The recurrent syndrome transformer AlphaQubit subsequently established state-of-the-art decoding accuracy on experimental Sycamore surface-code data and on realistic simulations through distance 11~\cite{bausch2024alphaqubit}; its successor AlphaQubit 2 extended high-accuracy neural decoding to substantially larger surface and color codes and demonstrated real-time variants on commercial accelerators~\cite{senior2025alphaqubit2}. 
However, in contrast to previous approaches, here we aggregate information spatially with an MLP and attend temporally with a transformer, instead of deploying the transformer to attend over spatial information and implement recurrence over time by construction. Moreover, this architecture, can inherently learn spatial correlations and features rather than relying on translational invariance, which might not be present due to transport details, crosstalk and gate errors which could vary across the ion trap.\\
We also highlight that, in contrast to the results in \cite{bausch2024alphaqubit, senior2025alphaqubit2}, both our ML models transfer from simulation training to hardware evaluation zero-shot, without hardware fine-tuning at all.

Both models were trained on data generated by the Helios-1 emulator to reproduce the noise of the Helios-1 QPU up-scaled by a factor of four for more efficient sampling of rare errors. For each distance and logical basis, the training pool contained up to $4.18\times10^8$ shots (the actual number of shots used depended on the specific model and data availability at the time of training, as specified in Tables \ref{tab:CNN_training} and \ref{tab:transformer_training}). A curriculum increased the number of syndrome-extraction rounds through $r=2,4,\ldots,18$, allocating $2\times10^6$ shots to each stage, before continuing at $r=20$. Both decoders were evaluated on the same unscaled multi-patch emulator data and hardware data, pooling both initial-state signs within each basis. The emulator test sets contained 121,200, 1,236,400, and 1,280,000 shots per basis at $d=3,5,7$, respectively; the hardware test sets contained 15,840, 21,600, and 29,000 shots in total across the three bases. Table~\ref{tab:ml_decoder_ler} reports the CNN and RTT emulator results for each basis and distance, combined LERs, and suppression factors. For each code distance, each architecture uses three basis-specific models, decoding the logical Pauli $X$, $Y$, and $Z$ eigenstates $\{\ket{\pm}, \ket{i\pm}, \ket{0},\ket{1}\}$, respectively.
The round-token transformer also used pre-trained models on basis $X$ and $Z$ to fine-tune for the decoding of the logical Pauli $Y$ basis states.

\begin{figure*}
    \centering
    \includegraphics[width=\textwidth]{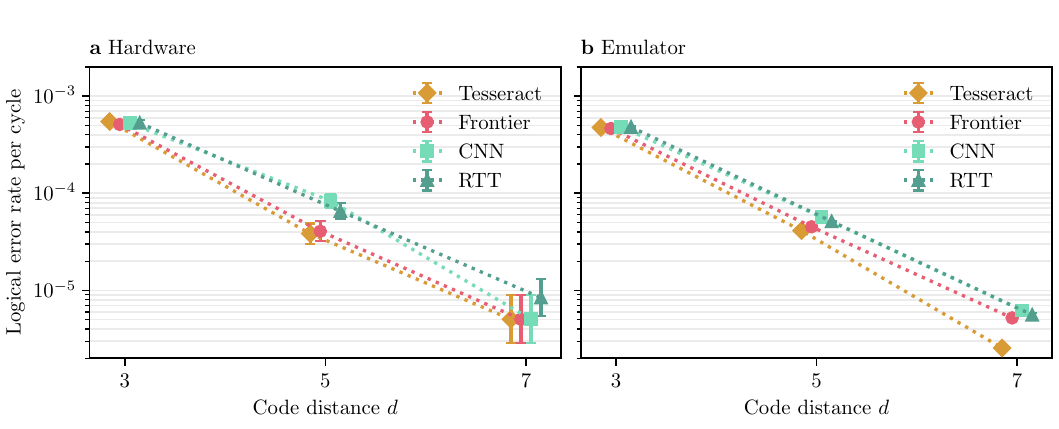}
    \caption{\textbf{Decoder logical error rates on hardware and emulator data.} (a) Hardware results for Tesseract, Frontier, CNN, and RTT, computed as $(n_F/n)/(r+1/2)$, where $n_F$ is the number of logical failures, $n$ is the total number of shots, and $r=20$ is the number of syndrome-extraction rounds. (b) Multi-patch emulator results for the same four decoders from Table~\ref{tab:ml_decoder_ler}.
    Emulator rates use the same normalization $(n_F/n)/20.5$ as the hardware results. Error bars are one-sigma Wilson intervals ($z=1$, nominal coverage $68.27\%$), normalized by the same factor of 20.5. Points are offset horizontally for visibility; dotted lines guide the eye.}
    \label{fig:hardware_emulator_decoder_ler}
\end{figure*}

Figure~\ref{fig:hardware_emulator_decoder_ler} compares the distance scaling of the hardware and emulator results reported in Tables~\ref{tab:shot_stats} and~\ref{tab:ml_decoder_ler}. We estimate finite-shot uncertainty in the suppression factors using a Wilson-weighted log-linear fit, $\log(\mathrm{LER}(d))=a+bd$, with $\mathrm{LER}=(n_F/n)/20.5$. Failure and shot counts are pooled before computing LERs and one-sigma Wilson intervals. The absolute fit uncertainty at each distance is $\log(\mathrm{LER}_{\rm upper})-\log(\mathrm{LER})$, using the upper one-sigma Wilson endpoint. The fitted suppression factor is $\Lambda=\exp(-2b)$, with one-sigma bounds $\exp[-2(b\pm\sigma_b)]$ from the absolute slope covariance. Adjacent-distance ratios and their intervals use the same fit restricted to the corresponding two distances. These intervals condition on the trained models and independent shots and do not quantify emulator mismatch or variation between training runs.

\begin{figure*}[t]
    \centering
    \includegraphics[width=\textwidth]{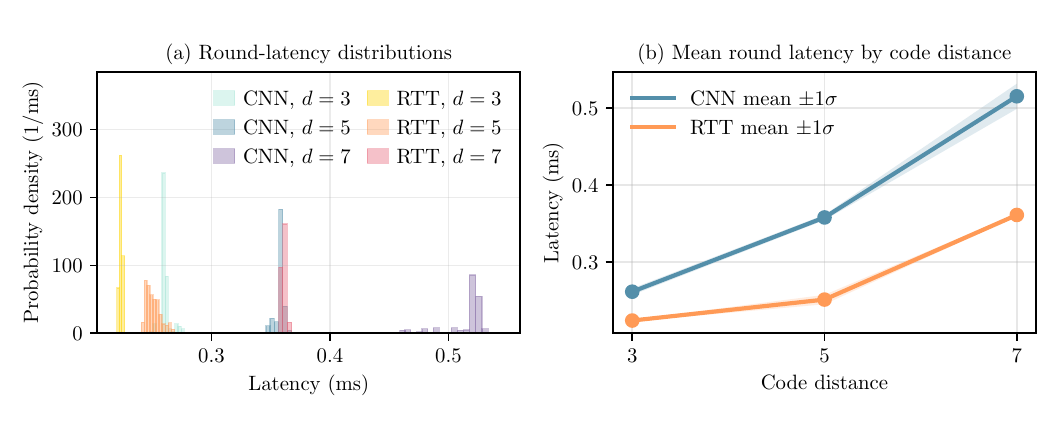}
    \caption{Per-round decoding latency for the CNN and RTT decoders. (a) Latency probability densities at code distances $d=3,5,7$, with cool colors for CNN and warm colors for RTT. (b) Mean latency versus code distance; shaded bands indicate one standard deviation of the latency samples around each mean. Lines, markers, and shaded bands use each model's middle shade from (a). Measurements use isolated GPU compute times with compilation and CUDA graphs, pooled over shots and steps.}
    \label{fig:rtt_cnn_round_latency}
\end{figure*}

Figure~\ref{fig:rtt_cnn_round_latency} compares the per-round latency distributions and their means for the CNN and RTT decoders at code distances $d=3,5,7$. Both models have mean per-round latencies well below 1 ms at the distances considered, with the RTT a factor $\approx 1.5$ faster than the CNN architecture at each distance. Measurements use an NVIDIA H100 80GB GPU with torch compilation, optimization of the generated cache, and a CUDA graph implementation. Decoder-specific optimizations could further reduce the latency.

\begin{table*}
    \centering
    \setlength{\tabcolsep}{5pt}
    \caption{Per-cycle logical error rates (LER) for the CNN, RTT, Frontier, and Tesseract decoders on multi-patch emulator data. The evaluation pools both initial-state signs, with 121,200, 1,236,400, and 1,280,000 shots per basis at $d=3,5,7$, respectively. Per-cycle LERs use $r+1/2=20.5$ for $r=20$ syndrome-extraction rounds, with combined results pooled across the three bases. The corresponding error suppression rate $\Lambda$ is computed as $\Lambda_{d} = \text{LER}(d-2)/\text{LER}(d)$. LER uncertainties are one-sigma Wilson intervals ($z=1$, nominal coverage $68.27\%$), transformed from circuit failure probability to per-cycle LER. The overall fitted suppression factors are $\Lambda=8.7^{+0.2}_{-0.2}$ for CNN, $\Lambda=9.3^{+0.2}_{-0.2}$ for RTT, $\Lambda=9.9^{+0.2}_{-0.2}$ for Frontier, and $\Lambda=12.1^{+0.3}_{-0.3}$ for Tesseract. Suppression-factor uncertainties are one-sigma log-fit intervals as described above; adjacent-distance ratio intervals use two-distance log fits.}
    \label{tab:ml_decoder_ler}
    \begin{tabular}{c ccccc}
        \toprule
        \multirow{2}{*}{$d$}
        & \multicolumn{4}{c}{LER}
        & \multirow{2}{*}{$\Lambda_{d-2 \rightarrow d}$} \\
        \cmidrule(lr){2-5}
        & $X$ & $Y$ & $Z$ & Combined & \\
        \midrule
        \multicolumn{6}{c}{CNN} \\
        \midrule
        3 & $4.2^{+0.1}_{-0.1}\!\times\!10^{-4}$ & $6.6^{+0.2}_{-0.2}\!\times\!10^{-4}$ & $3.8^{+0.1}_{-0.1}\!\times\!10^{-4}$ & $4.84^{+0.08}_{-0.08}\!\times\!10^{-4}$ & {--} \\
        5 & $4.5^{+0.1}_{-0.1}\!\times\!10^{-5}$ & $8.9^{+0.2}_{-0.2}\!\times\!10^{-5}$ & $3.6^{+0.1}_{-0.1}\!\times\!10^{-5}$ & $5.68^{+0.09}_{-0.09}\!\times\!10^{-5}$ & $8.5^{+0.2}_{-0.2}$ \\
        7 & $4.5^{+0.4}_{-0.4}\!\times\!10^{-6}$ & $1.05^{+0.07}_{-0.06}\!\times\!10^{-5}$ & $3.6^{+0.4}_{-0.4}\!\times\!10^{-6}$ & $6.2^{+0.3}_{-0.3}\!\times\!10^{-6}$ & $9.2^{+0.5}_{-0.4}$ \\
        \midrule
        \multicolumn{6}{c}{RTT} \\
        \midrule

        3
        & $4.2^{+0.1}_{-0.1}\!\times\!10^{-4}$
        & $6.5^{+0.2}_{-0.2}\!\times\!10^{-4}$
        & $3.7^{+0.1}_{-0.1}\!\times\!10^{-4}$
        & $4.81^{+0.08}_{-0.08}\!\times\!10^{-4}$
        & {--}  \\

        5
        & $4.7^{+0.1}_{-0.1}\!\times\!10^{-5}$
        & $7.2^{+0.2}_{-0.2}\!\times\!10^{-5}$
        & $3.5^{+0.1}_{-0.1}\!\times\!10^{-5}$
        & $5.14^{+0.08}_{-0.08}\!\times\!10^{-5}$
        & $9.4^{+0.2}_{-0.2}$ \\

        7
        & $4.3^{+0.4}_{-0.4}\!\times\!10^{-6}$
        & $9.0^{+0.6}_{-0.6}\!\times\!10^{-6}$
        & $3.6^{+0.4}_{-0.4}\!\times\!10^{-6}$
        & $5.6^{+0.3}_{-0.3}\!\times\!10^{-6}$
        & $9.1^{+0.5}_{-0.4}$ \\
        \midrule
        \multicolumn{6}{c}{Frontier} \\
        \midrule
        3 & $4.2^{+0.1}_{-0.1}\!\times\!10^{-4}$ & $6.3^{+0.2}_{-0.2}\!\times\!10^{-4}$ & $3.5^{+0.1}_{-0.1}\!\times\!10^{-4}$ & $4.65^{+0.08}_{-0.08}\!\times\!10^{-4}$ & {--} \\
        5 & $4.2^{+0.1}_{-0.1}\!\times\!10^{-5}$ & $6.4^{+0.2}_{-0.2}\!\times\!10^{-5}$ & $2.9^{+0.1}_{-0.1}\!\times\!10^{-5}$ & $4.52^{+0.08}_{-0.08}\!\times\!10^{-5}$ & $10.3^{+0.2}_{-0.2}$ \\
        7 & $4.2^{+0.4}_{-0.4}\!\times\!10^{-6}$ & $7.9^{+0.6}_{-0.5}\!\times\!10^{-6}$ & $3.5^{+0.4}_{-0.3}\!\times\!10^{-6}$ & $5.2^{+0.3}_{-0.3}\!\times\!10^{-6}$ & $8.7^{+0.5}_{-0.4}$ \\
        \midrule
        \multicolumn{6}{c}{Tesseract} \\
        \midrule
        3 & $4.3^{+0.1}_{-0.1}\!\times\!10^{-4}$ & $6.5^{+0.2}_{-0.2}\!\times\!10^{-4}$ & $3.4^{+0.1}_{-0.1}\!\times\!10^{-4}$ & $4.75^{+0.08}_{-0.08}\!\times\!10^{-4}$ & {--} \\
        5 & $3.8^{+0.1}_{-0.1}\!\times\!10^{-5}$ & $5.8^{+0.2}_{-0.1}\!\times\!10^{-5}$ & $2.8^{+0.1}_{-0.1}\!\times\!10^{-5}$ & $4.12^{+0.07}_{-0.07}\!\times\!10^{-5}$ & $11.5^{+0.3}_{-0.3}$ \\
        7 & $2.0^{+0.3}_{-0.3}\!\times\!10^{-6}$ & $3.9^{+0.4}_{-0.4}\!\times\!10^{-6}$ & $1.7^{+0.3}_{-0.2}\!\times\!10^{-6}$ & $2.6^{+0.2}_{-0.2}\!\times\!10^{-6}$ & $16^{+1}_{-1}$ \\
        \bottomrule
    \end{tabular}
\end{table*}

\begin{figure*}[t]
    \centering
    \includegraphics[width=\textwidth]{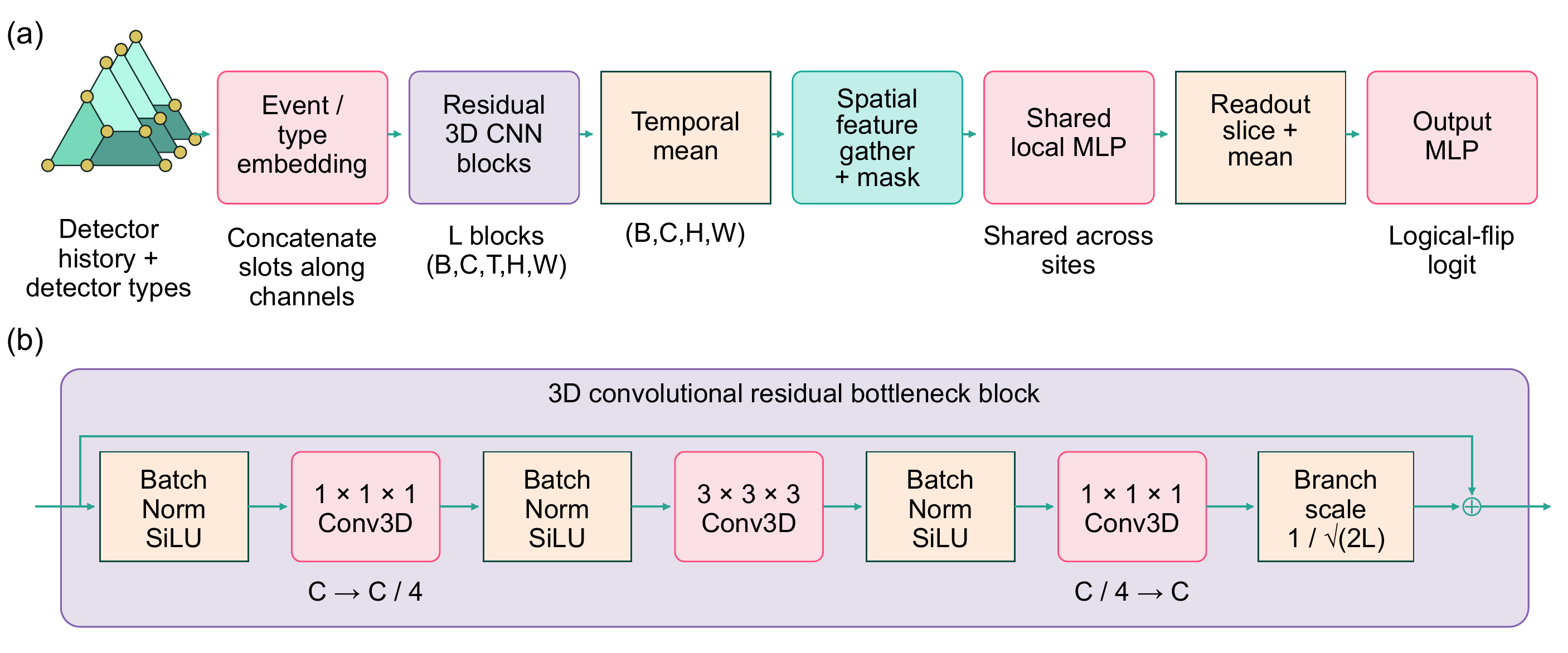}
    \caption{Architecture of the CNN decoder. (a) The available detector history and detector-type labels are embedded and processed by residual 3D convolutional blocks. The readout averages features over time, selects a spatial readout slice after a shared local MLP, and averages over the slice. A final MLP maps the pooled representation to a logical-flip logit. Projective-initialization flags, when present, are embedded and prepended as an additional time slice. (b) A pre-activation residual bottleneck block, with channel reduction, local space--time convolution, channel expansion, and residual-branch scaling. Here $C$ is the backbone channel width and $L$ is the number of residual blocks.}
    \label{fig:CNN3Ddecoder}
\end{figure*}

\paragraph{3D CNN}
Figure~\ref{fig:CNN3Ddecoder} summarizes the architecture of the CNN decoder.
The input to the CNN is the available detector history arranged using a general lattice-coordinate construction. Consider a topological code whose check locations lie on an $N$-dimensional Euclidean lattice. Choose primitive translation vectors $\mathbf{a}_1,\ldots,\mathbf{a}_N$ and write each check position as $\mathbf{r}=\mathbf{r}_0+\sum_{i=1}^{N}n_i\mathbf{a}_i+\boldsymbol{\rho}_\alpha$, where $n_i\in\mathbb{Z}$ and $\boldsymbol{\rho}_\alpha$ specifies a check within the unit cell. The dual (reciprocal) basis $\mathbf{b}_1,\ldots,\mathbf{b}_N$  gives the array coordinates through
\begin{equation}
    \mathbf{b}_i\cdot\mathbf{a}_j=\delta_{ij},
    \qquad n_i=\mathbf{b}_i\cdot(\mathbf{r}-\mathbf{r}_0-\boldsymbol{\rho}_\alpha),
    \qquad i,j=1,\ldots,N.
\end{equation}
After shifting the indices to be nonnegative, the finite patch fits in an $N$-dimensional rectangular array, with unused entries padded and distinct checks within each unit cell assigned separate channel slots. Local check neighbourhoods occupy bounded coordinate offsets, allowing finite convolutional kernels to capture spatial correlations. Stacking syndrome-extraction rounds supplies an additional axis, so the construction extends to an $(N+1)$-dimensional CNN acting on $N$ spatial coordinates and time. Including batch dimension $B$ and learned feature channel dimension $C$, the tensor has shape $(B,C,T,L_1,\ldots,L_N)$, where $L_i$ is the array extent along spatial axis $i$ and $T$ is the temporal dimension.

\begin{figure*}[t]
    \centering
    \includegraphics[width=\textwidth]{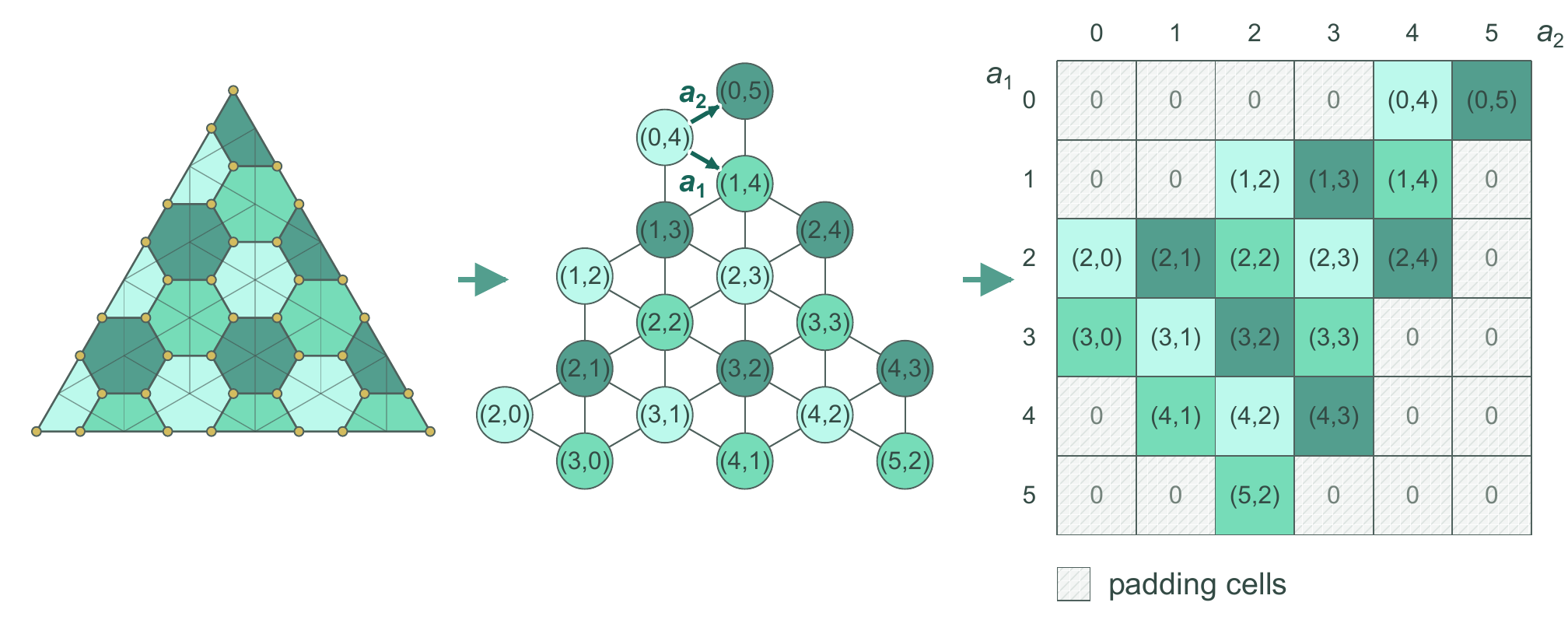}
    \caption{
    Lattice-coordinate construction of the two-dimensional array representation used by the CNN decoder, illustrated for triangular color-code patches. Stabilizer faces are assigned integer coordinates in a primitive basis of the triangular dual lattice, obtained by projection onto the reciprocal basis. Patches of different code distances use the same basis and are padded to rectangular arrays. 
    }
    \label{fig:2d_array_mapping}
\end{figure*}

For the honeycomb color code studied here, $N=2$ and stabilizer faces correspond to vertices of the triangular dual lattice, whose primitive translations provide $\mathbf{a}_1,\mathbf{a}_2$ (Fig.~\ref{fig:2d_array_mapping}). This geometric dual lattice specifies the check connectivity; the reciprocal basis above extracts coordinates in its translation lattice. X- and Z-type checks on the same face share spatial indices and occupy separate channel slots. At each occupied slot, the binary detector outcome is combined with an integer label identifying the detector type, so that the learned embedding distinguishes an X-type event from a Z-type event as well as an event from a non-event. The slot embeddings are concatenated along the channel dimension. Projective-initialization flags, when present, are embedded independently, projected to the same channel width, and prepended as an additional time slice. The resulting tensor has shape $(B,C,T,H,W)$, where $B$ is the batch size, $C$ is the learned feature width, and $(T,H,W)$ describes the space--time syndrome volume.

Final detector events compare the stabilizer parity inferred from the final data-qubit measurements with the corresponding outcome from the last syndrome-extraction round, firing when these disagree. These events are appended as an additional time slice after the last syndrome-extraction round, at the same spatial coordinates and in the same channel slots as the corresponding stabilizer checks. Only checks compatible with the physical readout basis contribute final detectors; the other slots contain zero detector values and retain their static detector-type labels. The final slice uses the same learned embedding and residual convolutional blocks as the syndrome history, allowing its events to combine with preceding rounds, and is included in the temporal average used for readout. Thus, for $r$ syndrome-extraction rounds, the complete input has $T=r+1$ time slices, or $T=r+2$ when projective-initialization flags are prepended.

The intermediate numerical values computed by the CNN (the activations), form arrays of learned features indexed by lattice position and syndrome round. Each convolutional layer combines nearby values to capture progressively larger space--time patterns. Activations whose inputs are unaffected by a new syndrome round can therefore be stored and reused.

\begin{figure*}[t]
    \centering
    \includegraphics[width=\textwidth]{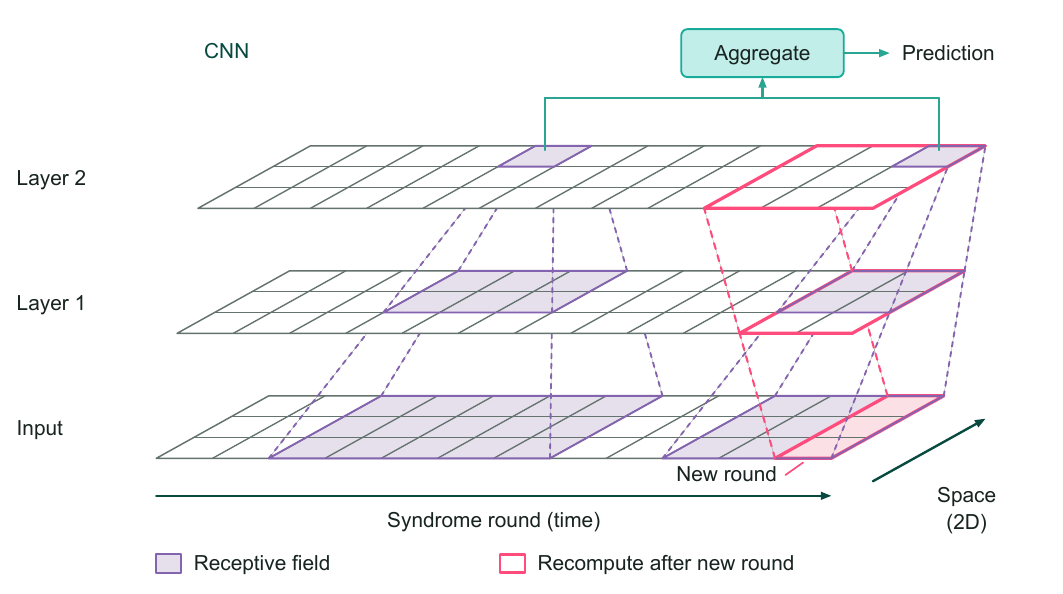}
    \caption{Causal-cone structure of the CNN during streaming inference, illustrated for two successive convolutional layers. Each plane represents a space--time feature volume, with the two spatial dimensions schematically projected onto one diagonal axis. Lavender regions show the receptive fields of two selected output features; dashed lines connect their boundaries across layers. The pink input slice denotes the newly available syndrome round, and pink outlines mark the temporal boundary region that must be computed or updated at each layer. This region extends one round further into the history with each convolutional layer. Earlier activations can be reused, and output features are aggregated to produce the decoder prediction.}
    \label{fig:CNN_causal_cones}
\end{figure*}

The backbone consists of $L$ residual bottleneck blocks. In each block, batch normalization and a SiLU activation precede a $1\times1\times1$ convolution that reduces the channel width by a factor of four. A same-padded $3\times3\times3$ convolution then mixes information between neighboring detector sites and adjacent rounds, after which a second $1\times1\times1$ convolution restores the original channel width. The result is added to the block input through a residual connection, with the residual branch scaled by $1/\sqrt{2L}$ to keep activation magnitudes stable as the network becomes deeper. Because every central convolution has unit stride and symmetric padding, the lattice dimensions are preserved throughout the network. Each block expands the receptive field by one site in every spatial and temporal direction, so successive blocks combine elementary local fault signatures into progressively longer space--time correlations while remaining fully parallel over the available volume. In streaming inference, part of this volume is recomputed after each new round, with symmetric padding at the current temporal boundary. Figure~\ref{fig:CNN_causal_cones} illustrates the resulting receptive fields and the temporal boundary regions that must be updated as each new syndrome round becomes available.

The readout reduces the feature tensor by averaging over one dimension, selecting a readout slice, and averaging over that slice. First, averaging along the time axis reduces the tensor from $(B,C,T,H,W)$ to $(B,C,H,W)$. A fixed spatial index map then gathers local groups of feature vectors, which are concatenated and processed by a shared two-layer MLP; masking accommodates incomplete groups at patch boundaries. A second fixed index selection extracts the readout slice from these transformed representations. Averaging over the sites in this slice yields a vector of fixed feature dimension, independent of the number of selected sites. A final two-layer MLP maps this vector to a logical-flip logit. The spatial index map and slice selection are specified by the patch's readout layout.

\begin{table*}
\centering
\setlength{\tabcolsep}{6pt}
\caption{Hyperparameters of the CNN decoder at each code distance. Each binary detector event and its detector-type label are embedded into 512 features. The two detector slots at each lattice position are concatenated to give the 1024 channels processed by the residual backbone. Each pre-activation residual block reduces this width to 256 channels, applies the stated same-padded convolution, and restores the channel width. The shared local MLP before slice selection and the final two-layer MLP both use this width, with final hidden width $d_{\mathrm{readout}}$.}
\begin{tabular}{|c|c|c|c|c|c|c|c|}
\hline
\shortstack{Code\\distance} &
\shortstack{Embedding\\width} &
Channels &
\shortstack{Residual\\blocks} &
\shortstack{Bottleneck\\width} &
\shortstack{Convolution\\kernel} &
$d_{\mathrm{readout}}$ &
\shortstack{Total parameters\\($\times 10^{6}$)} \\
\hline
3 & \multirow{3}{*}{512} & \multirow{3}{*}{1024} & 4 & \multirow{3}{*}{256} & \multirow{3}{*}{$3\!\times\!3\!\times\!3$} & \multirow{3}{*}{1024} & 14.97 \\
\cline{1-1}\cline{4-4}\cline{8-8}
5 & & & 6 & & & & 19.56 \\
\cline{1-1}\cline{4-4}\cline{8-8}
7 & & & 8 & & & & 24.16 \\
\hline
\end{tabular}
\label{tab:CNN_3D_params}
\end{table*}

\emph{Training}: For each distance and logical basis, the CNN training pool contained approximately $4.18\times10^8$ shots. A curriculum increased the number of syndrome-extraction rounds through $r=2,4,\ldots,18$, allocating $2\times10^6$ shots to each stage, before continuing at $r=20$. An epoch consisted of $10^6$ training examples, drawn from successive portions of the stored training data and shuffled within the epoch. Validation used a separate $4\times10^6$ shots from the unscaled emulator noise model every two epochs at $r=20$ with batch size 4096. Checkpoints were selected by minimum validation logical-failure fraction; early-stopping patience was ten validation checks. Training used noise scale 4, while emulator validation and test used scale 1.

A separate CNN was trained for each pair of code distance $d\in\{3,5,7\}$ and logical basis $X,Y,Z$, using the parameters in Table~\ref{tab:CNN_3D_params}. The hardware results in Table~\ref{tab:shot_stats} and emulator results in Table~\ref{tab:ml_decoder_ler} use the same nine validation-selected checkpoints listed in Table~\ref{tab:CNN_training}. The inputs were the two-dimensional arrays of stabilizer detection events and the projective-initialization flags, with the observable-flip bit as the binary target.

The optimizer used was AdamW with $(\beta_1,\beta_2)=(0.9,0.999)$, $\epsilon=10^{-8}$, and weight decay $10^{-5}$; bias and normalization parameters were excluded from weight decay, and AMSGrad was disabled. The batch size was 1024 at $d=3,5$ and 512 at $d=7$, with initial learning rates $2\times10^{-4}$ and $10^{-4}$, respectively. We observed that reducing the batch size reduced the number of training shots needed to reach convergence. No gradient clipping was applied. The loss was the batch mean of $0.84(1-p_t)^2[-\log(p_t)]$, where $p_t$ is the predicted probability of the correct binary label. Thus the focal-loss coefficient 0.84 multiplies both classes, rather than assigning complementary weights to the two labels. A cosine-annealing schedule updated the learning rate after every optimizer step, with $T_{\max}=294{,}000$ steps and $\eta_{\min}=5\times10^{-6}$. The learning rate was clamped to $\eta_{\min}$ past $T_{\max}$ steps.

Training used seed 42, deterministic trainer settings, bfloat16 mixed precision, and the highest float32 matrix-multiplication precision setting. Each run used one GPU at a time: an NVIDIA H100 80\,GB for the initial training, followed by resuming training on an NVIDIA H200. The recorded software environment was Python 3.12.11, PyTorch 2.8.0 with CUDA 12.8, cuDNN 9.10.2, NumPy 2.2.6. Across the nine selected models, the cumulative cost was $2.254\times10^9$ processed training shots and 229.03 single-GPU hours. The reported times measure elapsed trainer-fit time, including validation and in-fit overhead, and exclude offline data generation. They sum time on both GPU types and should not be interpreted as measurements on a single common device.

\begin{table*}
\centering
\caption{Training resources at the selected CNN checkpoints used for the reported evaluation. Training began on a H100 GPU and later resumed on a H200 GPU. The time after resuming training is shown in addition to the total training time.}
\label{tab:CNN_training}
\begin{tabular}{ccrrr}
\toprule
$d$ & Basis & Shots ($10^6$) & Total time (h) & Final resume (h) \\
\midrule
3 & $X$ & 228 & 4.65 & 0.71 \\
3 & $Y$ & 238 & 4.84 & 0.44 \\
3 & $Z$ & 274 & 5.36 & 2.16 \\
5 & $X$ & 356 & 29.03 & 12.19 \\
5 & $Y$ & 294 & 24.53 & 5.55 \\
5 & $Z$ & 270 & 22.68 & 3.89 \\
7 & $X$ & 172 & 38.65 & 15.86 \\
7 & $Y$ & 198 & 46.24 & 8.64 \\
7 & $Z$ & 224 & 53.05 & 11.23 \\
\bottomrule
\end{tabular}
\end{table*}

\paragraph{Round-token transformer}
\begin{figure*}[t]
    \centering
    \includegraphics[width=\textwidth]{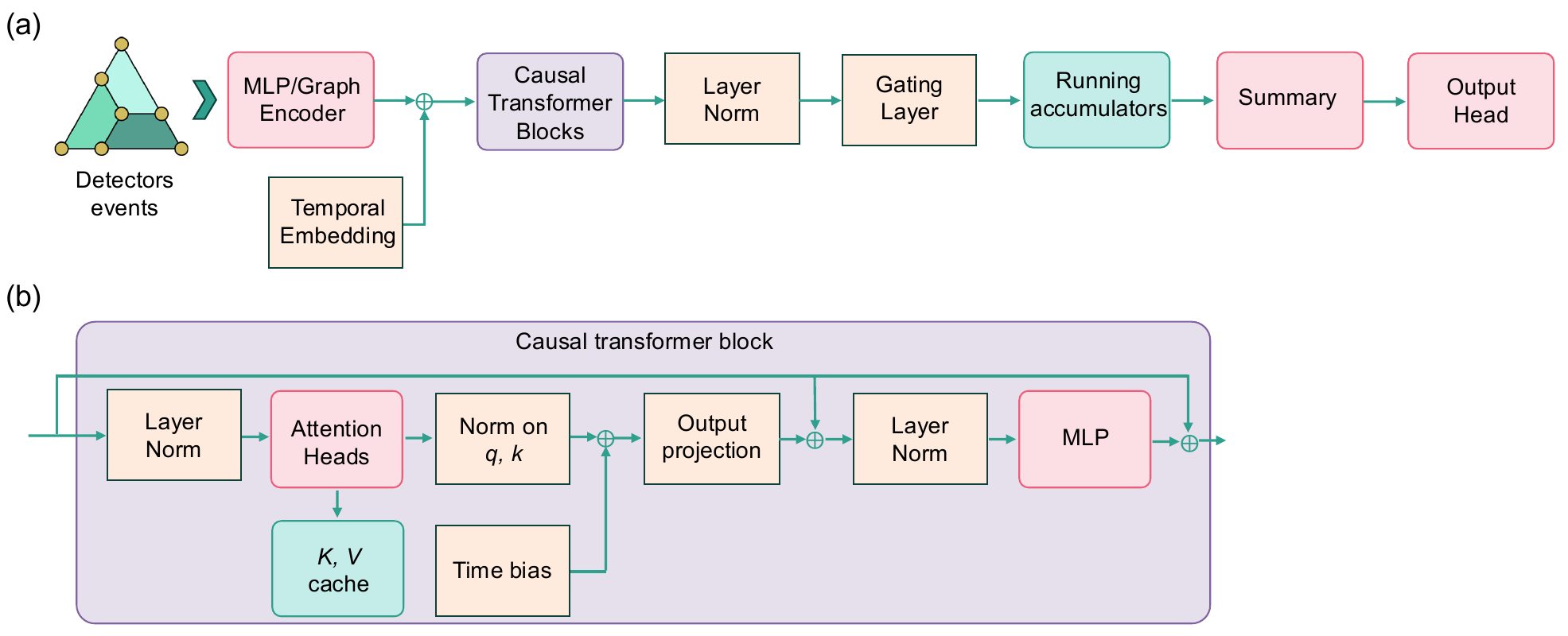}
    \caption{ Scheme of the architecture of the round-token transformer decoder. In (a) we represent the overall pipeline from syndrome extraction to logit output, in (b) the detailed structure of the causal transformer block. Additional description of details is provided in the main text.}
    \label{fig:MLdecoder}
\end{figure*}

At each syndrome extraction round of the memory experiment, we collect $X$- and $Z$-type detector events, which indicate that the corresponding stabilizer measurement parity differs from its expected value. 
For an $R$-round memory experiment, we get for example an array
$\bigl[X_{\mathrm{init}},X_1,Z_1,X_2,Z_2,\ldots,X_R,Z_R,X_{\mathrm{final}}\bigr]$, where each element is a vector of $P$ detector bits, one per stabilizer plaquette ($P=3,9,18$ for $d=3,5,7$, respectively). For each round $r$, the input of the decoder is
\begin{equation}
  f_r = \bigl[e_{r,1}^{x},\ldots,e_{r,P}^{x},
              e_{r,1}^{z},\ldots,e_{r,P}^{z},b_r,1-b_r\bigr]
\end{equation}
where the first $P$ elements are all the $X$-type detector events, the following $P$ are the $Z$-type ones, and the last two indicate whether it is a starting/ending event ($b_r=0/1$, respectively), to help discriminate the $Z$-type null events. All the elements $e_{r,i}^{X/Z}$ are binary detector events computed as the sum $\bmod\,2$ between two consecutive measurement outputs:
\begin{equation}
  e_{r,i}^{X/Z}
  = \bigl(m_{r-1,i}^{X/Z}+m_{r,i}^{X/Z}\bigr)\bmod 2,
\end{equation}
with $m_{r-1,i}^{X/Z}=0$ for $r=1$. \\

\emph{MLP encoder.} The MLP-based encoder processes this array using a small MLP, producing a vector of floating point elements which form a round's token:
\begin{equation}
  f_r \xrightarrow{\mathrm{Linear}} h_r
      \xrightarrow{\mathrm{GELU}} \widetilde{h}_r
      \xrightarrow{\mathrm{Linear}} t_r,
\end{equation}
or explicitly:
\begin{equation}
  t_r = W_2\operatorname{GELU}\!\bigl(W_1 f_r+b_1\bigr)+b_2,
\end{equation}
which for $d=5$ yields to the sizes $W_1=(512,20)$ and
$W_2=(384,512)$. Each hidden neuron of the MLP can combine information
from any $X$, $Z$ detector and flags, allowing the encoder to learn
spatial relationships and correlations between detector events within the same round. The output $t_r$ is a learned token vector. The same MLP weights process every step, so there is no rounds mixing within the encoder and no temporal correlation is analyzed at this level. The token's components are learned features, and spatial correlations are encoded and mainly caught at this stage. This construction avoids creating a single token for each detector, and allows to generalize to lower dimensionality via padding of the input of the neural network. \\
The token produced by the encoder is then modified by the addition of one more vector, namely a start embedding, providing information on how far the token's round represented is from the start:
\begin{equation}
  t_i^{\mathrm{encoded}}
  \longrightarrow t_i^{\mathrm{encoded}}\oplus\min(i,4).
\end{equation}
The values of this vector are clipped to 4 by choice, so that the bulk tokens are not intrinsically distinct from each other by construction, helping generalization across different number of rounds.\\
The tokens are then passed through multiple causal transformer blocks, processing only the token of the present round, masking the future ones if provided. Inside each block, the inputs are normalized, and then attended by multiple heads. Queries and keys are normalized, and the latest keys and values are cached at each round. The workflow is as follows:

\begin{enumerate}
  \item
    compute queries, keys and values ($Q,K,V$) for the new token;
  \item
    store current $K$ and $V$ alongside those from earlier tokens;
  \item
    use the new query to attend to all available keys and values,
    including the present ones, producing the attention weights
    \begin{equation}
      a_{ij}=\operatorname{softmax}_j\!\left(
        \frac{q_i\,k_j}{\sqrt{d_{\mathrm{head}}}}+b_{ij}
      \right)
    \end{equation}
    and the output is
    \begin{equation}
      o_r=\sum_{j=0}^{r}a_{rj}\,v_j;
    \end{equation}

  \item
    update the current token through the attention residual and
    feed-forward network.
\end{enumerate}

The tokens have shape $(B,S,D)$ while attention weights
$(B,N_{\mathrm{heads}},S,S)$, with $B$ batch size, $S$ sequence length (number of rounds = number of tokens per experiment), $D$ token dimension, $N_{\mathrm{heads}}$ the number of attention heads. During one streaming step, a token has shape $(B,1,D)$ and weights $(B,N_{\mathrm{heads}},1,n)$, where $n$ is the size of the available history (equal to round number).

Subsequently, a time bias is added to the attention scores in order to provide additional temporal information: a learned value
$\beta_h(j-t)$, for current token $t$ attending to source token $j$.

The output values of the attention heads are concatenated and projected out into a single vector, which is normalized and processed further by a narrow MLP with two residual connections, respectively. The resulting token is then further normalized and processed by a gating layer, which maps the token features to one value (between 0 and 1) used as a gate.
The gates determine the tokens' contributions to a normalized weighted mean (pooling). Each gate can therefore be calculated from its own token, without a \emph{softmax} over the whole sequence, allowing the model to produce an output logit even for incomplete sequences. The vectors output by the gate are pooled by running accumulators, which are fixed-size totals that let the model update its readout after each new token without pooling all previous tokens again. They include (i) the sum of the gates $g_j$ ($G=\sum_j g_j$), (ii) the gate-weighted sum of token vectors $h_j$
($A=\sum_j g_j h_j$), (iii) the plain sum of token vectors
($M=\sum_j h_j$), and the number of tokens $n$ processed so fare is also stored. From these totals a pooled quantity and the mean are computed:
\begin{equation}
  \mathrm{pooled}=\frac{A}{\max(G,10^{-6})},
  \qquad
  \mathrm{mean}=\frac{M}{n}.
\end{equation}
These two are concatenated and processed by the output head: a narrow MLP which produces a logit. From this number we calculate the normalized probability that the logical observable of the initial quantum state has flipped during the process.
The parameters used for the this architecture are listed in Table \ref{tab:Transformer_params}.\\

\emph{Training.} For every code distance and logical basis, the RTT training dataset contained $2\times10^{8}$ shots ($1.732\times10^{8}$ for $d=7$, $Z$ basis) with $r=20$ rounds, plus $2\times10^{6}$ shots for each round count $r=2,4,\ldots,18$. All of them had a noise intensity upscaled by a factor $4$. The last $2\times10^{6}$ shots of each $r=20$ dataset were equally split and used for validation and test, respectively, leaving $1.98\times10^{8}$ training shots ($1.712\times10^{8}$ for $d=7$, $Z$ basis). Each short-round dataset, instead, was split into $10^{6}$ training, $5\times10^{5}$ validation and $5\times10^{5}$ test shots. Final evaluation was performed on additional data without noise upscaling.
At every optimizer step one round count was chosen, either in strict rotation or from a fixed weighted schedule, and 4096 shot indices were drawn uniformly, with replacement, from that round count's training split. 
Each RTT model was trained in one or more stages. Every stage started from the weights of the selected checkpoint of the previous stage (when indicated), or from random initialization, with a fresh optimizer state, learning-rate schedule and weight average. 
Validation ran every 4000 optimizer steps (every 2000 in some short stages) and after the last step.
A separate RTT model was trained for each pair of code distance $d\in\{3,5,7\}$ and logical basis $X,Y,Z$. 
The batch size was 4096 at every distance and stage. For the loss we used the batch mean of the binary cross-entropy between the output logit and the observable-flip bit. Gradients were clipped to a global $\ell_2$ norm of 1.0 and, after the first 200 steps, additionally to three times an exponential moving average (decay 0.99) of recent gradient norms. The learning rate was updated after every optimizer step: it warmed up linearly from $\eta_{\max}/W$ to $\eta_{\max}$ over $W$ steps and then followed a cosine decay to $0.02\,\eta_{\max}$ at the last step of the stage, with $W$ 2\%, 3\%, 5\% or 10\% of the stage's steps, the number of steps is the stage's sample budget divided by the batch size, and $\eta_{\max}$ ranged from $5\times10^{-5}$ to $2\times10^{-4}$. 

The trainer used bfloat16 autocast, TF32 matrix multiplication and \texttt{torch.compile}. Each stage ran on a single NVIDIA H100 80 GB GPU. The packages used by the trainer were the following for all the trainings and evaluation: Python 3.12.13, PyTorch 2.8.0 with CUDA 12.8, cuDNN 9.10.2, NumPy 2.4.4 and Triton 3.4.0. The training of the decoders used a total of $1.601\times10^{9}$ distinct shots and processed $1.146\times10^{10}$ training examples in 49.73 single-GPU hours, counting every stage once, and only up to the checkpoint that was selected or passed on. Including the $X$-basis pre-training of the $d=7$, $Z$ basis, it used $1.895\times10^{9}$ distinct shots and processed $1.425\times10^{10}$ examples in 64.10 hours.

\begin{table*}
\centering
\caption{Training resources of the RTT models used for the hardware and emulator evaluations, counted up to the checkpoint that was selected or passed on to the next stage. The pool is the number of distinct shots in the dataset, used at least once: the bracketed value is the number of training examples processed, counting the repeated ones. Times are elapsed trainer hours on one NVIDIA H100 80 GB GPU. Rows marked with a superscript have been fine-tuned from a pre-trained model of a different basis: their pool and shots cover only the stages trained on that specific basis and distance, while the bracketed time adds the pre-training. The corresponding footnotes provide more details about it. The selected step is the one with the best validation out of all the training steps performed.}
\label{tab:transformer_training}
\setlength{\tabcolsep}{10pt}
\begin{tabular}{crrrrr}
\toprule
$d$ &  Basis & \quad Pool ($10^6$) & \quad Shots ($10^6$) & \quad Time (h) & \quad Selected step \\
\midrule
3 & $X$ & 207.0 & 18.6 (139) & 0.24 & 24,000 / 97,656 \\
3 & $Y$ & 198.0 & 197.8 (1,343) & 3.21 & 328,000 / 366,210  \\
3 & $Z$ & 198.0 & 197.1 (1,065) & 2.54 & 260,000 / 366,210 \\
5 & $X$ & 207.0 & 204.7 (967) & 3.81 & 236,000 / 244,140 \\
5 & $Y$$^{(a)}$ & 198.0 & 197.9 (1,458) & 4.62 (8.59) & 356,000 / 366,210 \\
5 & $Z$ & 198.0 & 197.6 (1,245) & 3.96 & 304,000 / 366,210 \\
7 & $X$ & 207.0 & 207.0 (1,851) & 10.85 & 420,000 / 854,492 \\
7 & $Y$$^{(b)}$ & 207.0 & 204.8 (1,278) & 8.20 (34.87) & 240,000 / 244,140 \\
7 & $Z$$^{(c)}$ & 176.2 & 176.1 (2,117) & 12.29 (26.66) & 172,000 / 558,492 \\
\bottomrule
\end{tabular}
\\[4pt]
\begin{minipage}{0.95\textwidth}\footnotesize
$^{(a)}$ Initialized from the selected $d=5$ $Z$ basis checkpoint. \\
$^{(b)}$ Initialized from the selected $d=7$ $Z$ basis checkpoint. \\
$^{(c)}$ Initialized from a $d=7$ $X$-basis model previously trained on an earlier dataset. The pre-training details enters the bracketed ones of this model. \\
\end{minipage}
\end{table*}

\begin{table*}
\centering
\setlength{\tabcolsep}{6pt}
\caption{Hyperparameters of the architecture of the round-token transformer decoder at each code distance. A two-layer MLP of hidden width $d_{\mathrm{enc. , MLP}}$ compresses the detector bits of one round into a token of the given length. That sequence passes through the stated number of pre-norm transformer blocks, each with the given number of attention heads; head width is the token length divided by the head count, and $d_{\mathrm{MLP}}$ is the hidden width of the feed-forward sub-layer, fixed at four times the token length.}
\begin{tabular}{|c|c|c|c|c|c|c|c|}
\hline
\shortstack{Code\\distance} &
\shortstack{Token\\length} &
\shortstack{Transformer\\blocks} &
\shortstack{Attention\\heads} &
\shortstack{Head\\width} &
$d_{\mathrm{MLP}}$ &
$d_{\mathrm{enc.\,MLP}}$ &
\shortstack{Total parameters\\($\times 10^{6}$)} \\
\hline
3 & 256 & 8 & 8 & 32 & 1024 & 512 & 6.79 \\
\hline
5 & 384 & 8 & 6 & 64 & 1536 & 512 & 15.15 \\
\hline
7 & 512 & 10 & 8 & 64 & 2048 & 1024 & 33.40 \\
\hline
\end{tabular}
\label{tab:Transformer_params}
\end{table*}

\end{document}